\documentclass[usenatbib,useAMS,letterpaper]{mn2e}
\usepackage{graphicx,times}

\title[X-ray selected clusters in the PS1 MDS] {X-ray selected galaxy clusters in the Pan-STARRS Medium-Deep Survey\thanks{Some of the data presented herein were obtained at the W.M. Keck Observatory, which is operated as a scientific partnership among the California Institute of Technology, the University of California and the National Aeronautics and Space Administration. The Observatory was made possible by the generous financial support of the W.M. Keck Foundation. }}

\author[Ebeling et al.] {\parbox{\textwidth}{H.\ Ebeling$^1$,
 A.C.\ Edge$^2$, W.S.\ Burgett$^1$, K.C.\ Chambers$^1$, K.W.\ Hodapp$^1$, M.E.\ Huber$^1$, N.\ Kaiser$^1$, P.A.\ Price$^3$, J.L.\ Tonry$^1$}\vspace{4mm}\\ $^1$Institute for Astronomy, University of
  Hawaii, 2680 Woodlawn Drive, Honolulu, HI 96822, USA\\ $^2$ Institute for Computational Cosmology, Department of Physics, Durham University, Durham, DH1 3LE, UK\\ $^3$Princeton University Observatory, 4 Ivy Lane, Peyton Hall, Princeton University, Princeton, NJ 08544, USA}

\date{MNRAS, April 2013}

\begin{document}

\label{firstpage}

\maketitle

\begin{abstract}
We present the results of a pilot study for the extended MACS survey (eMACS), a comprehensive search for distant, X-ray luminous galaxy clusters at $z>0.5$.  Our pilot study applies the eMACS concept to the 71 deg$^2$ area extended by the ten fields of the Pan-STARRS1 (PS1) Medium Deep Survey (MDS). Candidate clusters are identified by visual inspection of PS1 images in the g,r, i, and z bands in a $5\times5$ arcmin$^2$ region around X-ray sources detected in the ROSAT All-Sky Survey (RASS).  To test and optimize the eMACS X-ray selection criteria, our pilot study uses the largest possible RASS database, i.e.,  all RASS sources listed in the Bright and Faint Source Catalogs (BSC and FSC) that fall within the MDS footprint. We apply no additional constraints regarding X-ray flux, spectral hardness ratio, or photon statistics and lower the redshift threshold to $z>0.3$ to extend the probed luminosity range to poorer systems. Scrutiny of PS1/MDS images for 41 BSC and 200 FSC sources combined with dedicated spectroscopic follow-up observations results in a sample of  11 clusters with estimated or spectroscopic redshifts of $z>0.3$. 

In order to assess and quantify the degree of point source contamination of the observed RASS fluxes, we examine archival Chandra data obtained in targeted and serendipitous observations of six of the 11 clusters found. As expected, the diffuse emission from all six systems is contaminated by point sources to some degree, and for half of them AGN emission dominates. X-ray follow-up observations will thus be crucial in order to establish robust cluster luminosities for eMACS clusters.

Although the small number of distant X-ray luminous clusters in the MDS does not allow us to make firm predictions for the over 20,000 deg$^2$ of extragalactic sky covered by eMACS, the identification of two extremely promising eMACS cluster candidates at $z\ga 0.6$ (both yet to be observed with Chandra)  in such a small solid angle is encouraging. Representing a tremendous gain over the presently known two dozen such systems from X-ray, optical, and SZ cluster surveys combined, the sample of over 100 extremely massive clusters at $z{>}0.5$ expected from eMACS would be invaluable for the identification of the most powerful gravitational lenses in the Universe, as well as for in-depth and statistical studies of the physical properties of the most massive galaxy clusters out to $z\sim 1$.
\end{abstract}


\section{Introduction}

Massive galaxy clusters ($M{\sim}10^{15}$ M$_\odot$) play a central role in extragalactic astronomy. Containing vast amounts of dark and luminous matter, they are rewarding targets for studies of, e.g., galaxy evolution in dense environments, intra-cluster gas dynamics, or the properties of dark matter (e.g., Edge et al.\ 2002; Markevitch \& Vikhlinin 2007; Brada\v{c} et al.\ 2008; Ma et al.\ 2008, 2010; Korngut et al.\ 2011; Ma \& Ebeling 2011; Oguri et al.\ 2012;  Owers et al.\ 2012). Thanks to the very high mass surface density of their cores, massive clusters also represent extremely powerful gravitational lenses that allow the detection and study of faint and distant galaxies out to redshifts far beyond the reach of man-made telescopes (e.g., Limousin et al.\ 2007; Ebeling et al.\ 2009; Richard et al.\ 2011; Coe et al.\ 2012). In addition, statistical samples of massive clusters constitute highly sensitive probes of cosmological parameters already at $z\sim 0.3$ (e.g., Allen et al.\ 2008; Mantz et al.\ 2010), much more so than low-mass clusters which provide comparable leverage only at much higher redshift.

It follows that the availability of well selected samples of massive clusters, in particular at intermediate to high redshifts, is vital to many key topics of astrophysical and cosmological research, and much progress has been made in this regard in recent years.
From X-ray detections listed in the {\it ROSAT}\/ All-Sky Survey (RASS) Bright Source Catalogue (BSC; Voges et al.\ 1999), the Massive Cluster Survey (MACS; Ebeling, Edge \& Henry 2001) compiled an X-ray selected sample of over 120 very X-ray luminous clusters at $z>0.3$ (Ebeling et al.\ 2007, 2010; Mann \& Ebeling 2012) within a solid angle of over 22,000 deg$^2$, defined by $-40^{\circ} \leq \delta \leq 80^{\circ}$, $|b| > 20^{\circ}$. Increasing the number of such clusters known by over a factor of 30 over previous samples, MACS has enabled countless in-depth as well as statistical studies of massive clusters and of the distant Universe behind them (e.g., Allen et al.\ 2008; Smith et al.\ 2009; Swinbank et al.\ 2010; Coe et al.\ 2013). More recently, surveys exploiting the Sunyaev-Zel'dovich (SZ) effect (Sunyaev \& Zel'dovich 1972) have begun to probe the same extreme area of  mass-redshift space, adding additional clusters out to $z{\sim} 1$ [{\it Planck}, South Pole Telescope (SPT)]. X-ray follow-up has proven critical though for SZ cluster surveys in order to eliminate false positives caused by complex correlated noise patterns as well as unrelated radio sources. Again, the RASS has been invaluable in this regard: ``SZ candidates with no detection at all in [the] RASS are almost certainly false" (Planck Collaboration 2013).

We here present results from a pilot study for the extended Massive Cluster Survey (eMACS), a new, very large-area cluster survey that combines X-ray selection with optical confirmation in its quest for the most massive clusters at $z\ge 0.5$.
In Sections~\ref{sec:rass} and \ref{sec:ps1} we briefly introduce the X-ray and optical databases underlying eMACS. Section~\ref{sec:emacs} provides an overview of the eMACS project, followed by a discussion of the importance of contamination from X-ray point sources in Section~\ref{sec:xps}. The design of our pilot study as well as our results are presented in Sections~\ref{sec:pilot} and \ref{sec:sample}, respectively. We examine available {\it Chandra}\/ data for the clusters in our sample in Section~\ref{sec:cxo}, briefly discuss the relevance of cluster velocity dispersions in Section~\ref{sec:veldisp}, and close with a summary of our findings and their implications for future work (Section~\ref{sec:discussion}). Throughout we adopt the concordance $\Lambda$CDM cosmology with $\Omega_M=0.3$, $\Omega_\lambda=0.7$, and $H_0=70$ km s$^{-1}$ Mpc$^{-1}$. 

\section{X-ray source catalogues from the RASS} \label{sec:rass}

X-ray sources detected in the RASS are catalogued in two separate lists, the Bright Source Catalogue (BSC) and the Faint Source Catalogue (FSC). The former contains almost 19,000 sources that meet the criteria LH${\ge} 15$ and $n_{\rm src}{\ge}15$ (LH is the likelihood of detection, and $n_{\rm src}$ is the number of net source photons); the latter comprises approximately 106,000 additional detections down to LH${=}7$ and $n_{\rm src}{=}6$.  Since exposure time is not constant in the RASS, but varies from less than 100 s (for some  3 per cent of the sky) to over 10,000 seconds in the immediate vicinity of the ecliptic poles, the criteria used to separate FSC and BSC sources do not correlate with source flux. As a result, intrinsically bright sources can be found in both catalogs, which makes the much larger FSC a database of enormous promise for surveys of distant, X-ray luminous clusters.

\section{Pan-STARRS} \label{sec:ps1}

\begin{table*}
\caption{\label{tab:mdfields}
Field centres and RASS source statistics for the 10 fields of the PS1 MDS.  Acronyms used: {\it XMM}-LSS =  {\it XMM}\/ Large-Scale Structure project; VVDS = VIMOS-VLT Deep Survey; CDFS = {\it Chandra}\/ Deep Field South; GOODS-S = Great Observatories Origins Deep Survey (South); COSMOS = Cosmological Evolution Survey; DEEP2 = Deep Extragalactic Evolutionary Probe (Part 2), ELAIS = European Large-Area {\it ISO}\/ Survey; SA22 = Special Area 22. The average exposure time in the RASS is listed together with the observed dispersion within a given field. For each field, the final two columns list the number of RASS sources in the BSC and FSC, respectively.}
\begin{tabular}{llccccc}
& & \multicolumn{2}{c}{MDS field centre} & \multicolumn{3}{c}{RASS statistics}\\
Name & Description & \multicolumn{2}{c}{R.A.\ (J2000) Dec}  & $t_{\rm exp}$ (s) & $N$(BSC) & $N$(FSC)\\ \hline
MD01  & XMM-LSS / VVDS        & 02 23 30 & $-$04 15 00   & $210\pm61$ & 3& 10\\
MD02  & CDFS/GOODS-S          & 03 32 24 & $-$27 48 00  & $\;\;54\pm 55$ & 0& \,\,\,4\\
MD03  & IFA/Lynx                         & 08 42 22 & $+$44 19 00  & $327\pm 95$ & 4& \,\,\,8\\
MD04  & COSMOS                       & 10 00 00 & $+$02 12 00  & $432\pm 5\;\;$ & 4 & 27\\
MD05  & Lockman Hole              & 10 47 40 & $+$58 05 00  & $\;\;365\pm 103$ & 7 & 22\\
MD06  & NGC\,4258                    & 12 20 00 & $+$47 07 00  & $400\pm 38$  & 7 & 27\\
MD07  & DEEP2 Field 1              & 14 14 49 & $+$53 05 00  & $663\pm 17$ & 5& 39\\
MD08  & ELAIS-N1                       & 16 11 09 & $+$54 57 00   & $1044\pm 103$ & 6& 50\\
MD09  & SA22 / VVDS                & 22 16 45 & $+$00 17 00   & $251\pm 25$ & 3 & \,\,\,4\\
MD10  & DEEP2 Field 3              & 23 29 15 & $-$00 26 00   & $332\pm 27$ & 2& \,\,\,9\\
\end{tabular}
\end{table*}

Pan-STARRS (Kaiser et al.\ 2002) is a wide-field imaging facility  on the summit of Haleakala (Hawai'i).  At present, it consists of Pan-STARRS1 (PS1), a single 1.8-m telescope with a 7 deg$^2$ field of view,  a gigapixel camera with on-chip guiding capabilities, and a $g_{\rm P1}$, $r_{\rm P1}$, $i_{\rm P1}$, $z_{\rm P1}$, $y_{\rm P1}$ filter set (Tonry et al.\ 2012). PS1 began operations in 2010 March and has since embarked on several survey programmes, two of which are of particular interest for extragalactic astronomy: the Medium Deep Survey (MDS), covering the 10 fields listed in Table~\ref{tab:mdfields} and reaching exposure times of several hours in each passband, and the  ``3$\pi$'' survey\footnote{A detailed description of the $3\pi$ survey is provided by Chambers et al.\ (in preparation).}, a 3-year survey of the entire sky visible from Hawaii.

\section{eMACS}  \label{sec:emacs}

The eMACS project aims to expand the MACS cluster survey to higher redshift and lower X-ray fluxes by combining the two large-area imaging data sets introduced in the preceding sections: the RASS and the PS1 ``3$\pi$" survey.

Our strategy for the identification of galaxy clusters at $z>0.5$ from these data sets is brute force: we select all X-ray sources listed in the  RASS BSC and FSC that fall within our study area, and then examine PS1 images in the $g_{\rm P1}$, $r_{\rm P1}$, $i_{\rm P1}$, and $z_{\rm P1}$ bands in a $5\times5$ arcmin$^2$ region around the X-ray source position.  Candidate clusters at intermediate to high redshift ($z\ga0.3$) are readily identifiable as pronounced overdensities of faint, red galaxies.  In order to prevent seemingly blank fields from erroneously being classified as potentially very distant clusters, we also query the NASA Extragalactic Database (NED) for known celestial objects within 2 arcmin radius of the respective X-ray source, a process that eliminates large numbers of active galactic nuclei (AGN) and quasi-stellar objects (QSOs). Further details of the eMACS cluster selection strategy are provided by Ebeling et al.\ (in preparation).

Clusters selected in the process described above are scrutinized again prior to inclusion in the  spectroscopic follow-up phase of eMACS. To limit such targeted follow-up to the most promising candidates without sacrificing completeness, only systems estimated to be at $z\ge 0.4$ are selected. While the primary goal of spectroscopic observations is the measurement of the cluster redshift, secondary goals include an assessment of the cluster velocity dispersion (we aim to secure redshifts of at least 10 cluster members) and the spectroscopic identification of potential AGN and QSOs in the cluster vicinity. 

\section{Point-source contamination}  \label{sec:xps}

With X-ray point sources outnumbering galaxy clusters 100:1 per solid angle, contamination from AGN and QSOs poses a problem for all X-ray cluster surveys. Although optical evidence of an overdensity of galaxies at the location of an X-ray source  strongly suggests a cluster ID, the possibility of severe contamination from point sources, or in fact of a misidentification, always exists.

A simple probabilistic argument can be made to illustrate the relevance of cluster mass in this context. The comoving space density of clusters of low X-ray luminosity ($L_{\rm X} \sim 5\times 10^{43}$ erg s$^{-1}$) at $z\ga 0.5$ is about $10^{-6}$ Mpc$^{-3}$ (e.g., Mullis et al.\ 2004), almost 10 times higher than that of very X-ray luminous AGN ($L_{\rm X} \sim 10^{45}$ erg s$^{-1}$) at $z\le 1.5$ (Miyaji, Hasinger \& Schmidt 2000).  Hence, if the X-ray luminosity implied by the redshift of a distant, but optically poor cluster of galaxies near a RASS source approaches or exceeds $10^{45}$ erg s$^{-1}$, we are likely dealing with either a chance coincidence or a blend of point-like and cluster emission. This picture changes dramatically for very rich and massive clusters whose space density at $z\ga 0.5$ is only a few $10^{-10}$ Mpc$^{-3}$, about two orders of magnitudes lower than that of comparably X-ray luminous QSOs. A spatial coincidence of a RASS X-ray source with an optically rich cluster at $z\ga 0.5$ can thus be taken as almost certainly physical in nature. A quantitative example of this argument is discussed in detail by Zenn \& Ebeling (2010).

Although our focus on extremely massive clusters thus mitigates the risk of misidentifications, two caveats remain. For one, a cluster's optical richness is only loosely correlated with X-ray luminosity or mass, and hence significant contamination from X-ray point sources can never be ruled out from RASS data alone. Secondly, highly evolved (i.e., fully relaxed) clusters can appear deceivingly poor in the optical waveband, in particular in shallow images. Spatial coincidences between a RASS source and a seemingly lone, giant, distant elliptical galaxy must thus not be dismissed but scrutinized further until the presence of a massive cool-core cluster can be firmly ruled out.

\section{An eMACS pilot study}  \label{sec:pilot}

Our pilot study explores the validity of the eMACS survey strategy by applying it to PS1 data for the ten fields of the MDS.  At a limiting magnitude of 26.3 (5$\sigma$) for the $i_{\rm P1}$ band, the MDS is significantly deeper than the projected limit of the $3\pi$ survey ($m_{i, {\rm P1, lim}} = 22.5$). Our pilot study thus does not aim to test the efficiency or reliability of optical cluster confirmations obtained by eMACS based on PS1-$3\pi$ images. Instead, we use MDS data to eliminate any uncertainties caused by the limited depth of the $3\pi$ survey's images, thus allowing us to assess whether the faintest RASS sources, comprising no more than a handful of X-ray photons, indeed constitute credible detections that can be used to identify massive galaxy clusters to redshifts approaching $z=1$. 

A grand total of 41 BSC and 200 FSC sources fall within 1.5 degrees (radius) of the MDS field centres listed in Table~\ref{tab:mdfields}. The average RASS exposure time varies greatly between MDS fields (from a mere 50 to over 1,000 s), and in fact even within a given field. Since the log$N$--log$S$ distribution, i.e., the number of RASS X-ray sources of a given flux, increases strongly with decreasing source flux (e.g., Ebeling et al.\ 1998; Voges et al.\ 1999), the total projected surface density of RASS sources varies dramatically too (Table~\ref{tab:mdfields}).

The results of applying the eMACS cluster selection strategy (outlined in Section~\ref{sec:emacs}) to  RASS sources in our study area are described in the following section.

\section{Results} \label{sec:sample}

We subject each of the 241 RASS sources that fall within the footprint of the PS1 MDS to the screening process described in Section~\ref{sec:emacs}. Since the depth of the MDS is easily sufficient to reveal massive clusters out to $z\sim 1$ as pronounced overdensities of galaxies of similar colour, the results of this screening process are binary in nature: a given RASS source is classified either as a massive cluster at a measured or estimated redshift in excess of $z{=}0.3$ --- or not. Although we record plausible non-cluster identifications for statistical purposes (see below), we are not concerned at all about classifying RASS sources as of ``unknown origin", since our goal is not the identification of optical counterparts to all RASS sources, but only the identification of extremely massive and distant clusters. 

\begin{figure}
\hspace*{-5mm}
\includegraphics[width=0.51\textwidth]{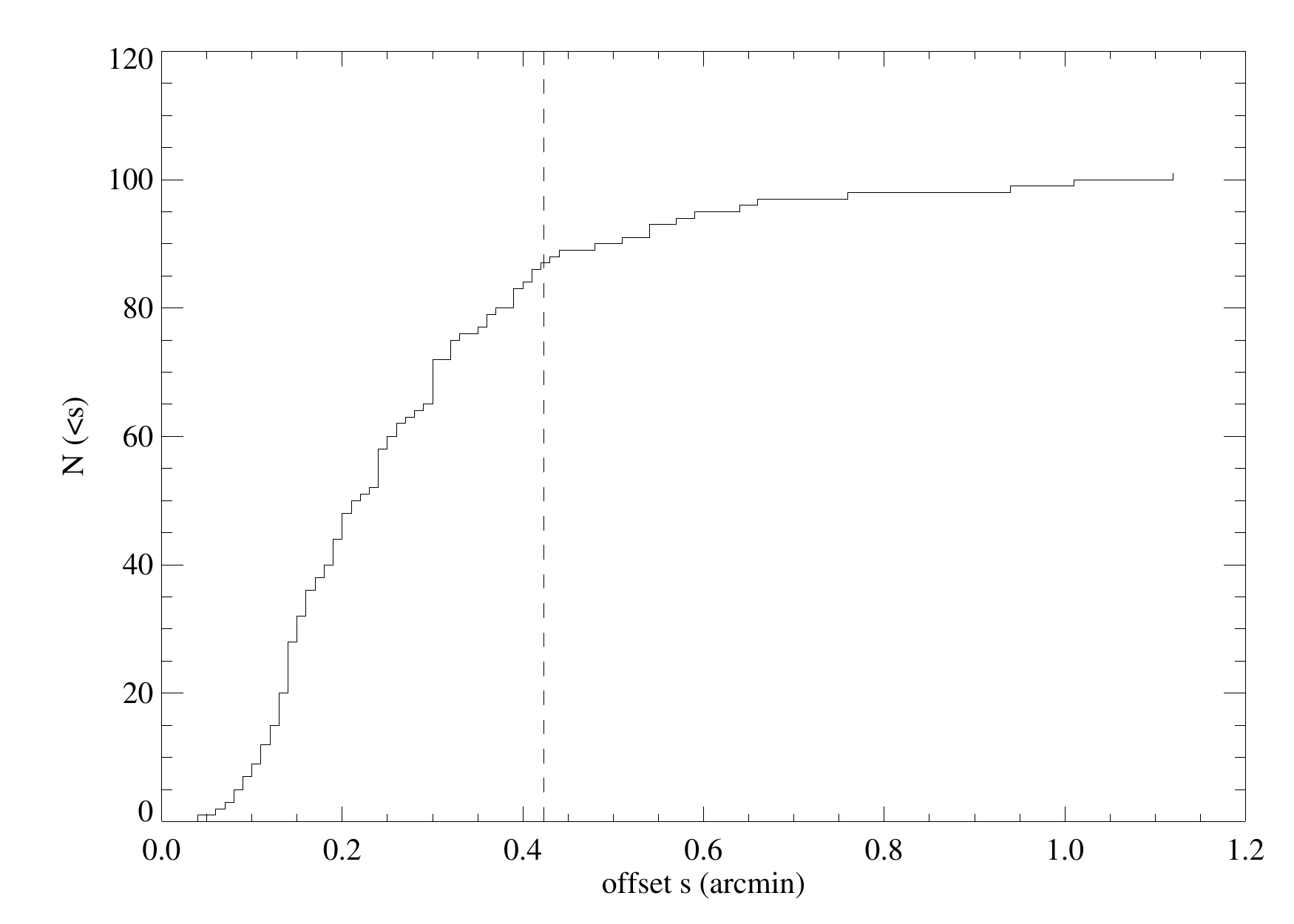}\hspace*{-3mm}
\caption{Cumulative distributions of the distance between a RASS source and the associated optical counterpart for the 96 BSC and FSC sources identified as QSO, AGN, or stars in our eMACS pilot project. The dashed vertical line shows the radius of a circle of the same area as a RASS pixel.\label{fig:xpsoff}}
\end{figure}

Based on entries listed in NED, we tentatively identify 96 of the 241 RASS sources within the PS1 MD fields as QSOs or AGN, and eight as bright stars. The distribution of offsets between the nominal RASS positions of these 96 X-ray sources and the location of the adopted optical counterpart is shown in Fig~\ref{fig:xpsoff}. The width of the distribution is in excellent agreement with the  pixel size of 45\arcsec, chosen for the RASS as an approximation of the size of the point-spread function of the {\it ROSAT} Position-Sensitive Proportional Counter (PSPC) in survey mode. 

27 BSC and FSC sources are classified as likely galaxy clusters. Of these, 12 have literature redshifts. For an additional six without cluster redshifts NED provides a spectroscopic redshift for the brightest cluster galaxy (BCG) which we adopt as the likely cluster redshift. Another five cluster candidates have estimated or photometric literature redshifts. To the remaining four we assign a crude redshift estimate based on their optical appearance. This first iteration of our  cluster compilation process results in a list of 11 clusters known or estimated to be at $z{>}0.3$  Entries in this list that do not have secure spectroscopic redshifts are selected for spectroscopic follow-up observations, described in more detail in the Appendix.

For all five clusters observed by us we find the measured redshift to confirm our estimate of $z{>}0.3$; three systems are found to lie at $z{\ga}0.6$. At the time of this writing, two clusters (with estimated redshifts of $z\sim 0.38$ and 0.65, respectively) still await spectroscopic confirmation. Basic properties of all 11 clusters are listed in Table~\ref{tab:mdclusters} where we assign eMACS names to all systems with $z{>}0.5$. In the field of one system (RXJ1613.7+5542) we find  a foreground broad-line AGN superimposed on the cluster core. Although we cannot quantify what fraction of the RASS flux can be attributed to this AGN, we expect it to contribute significantly and thus mark this cluster as contaminated in Table~\ref{tab:mdclusters}. The positions and redshifts of all individual galaxies successfully observed during our spectroscopic follow-up can be found in Table~\ref{tab:redshifts} where we also list  velocity dispersions for all five clusters.

\begin{table*}
\caption{\label{tab:mdclusters} Clusters at $z>0.3$ detected in the PS1 MD fields by our eMACS pilot study.  Clusters featuring redshifts in excess of $z=0.5$ are given an eMACS name, systems at lower redshift are referred to by their RASS ID. Coordinates mark the position of the RASS source. X-ray fluxes and luminosities in the 0.1--2.4 keV band are quoted in units of $10^{-12}$ erg s$^{-1}$ cm$^{-2}$ and  $10^{44}$ erg s$^{-1}$, respectively. Velocity dispersions ($\sigma$) are quoted in km s$^{-1}$ in the cluster rest-frame; values in parentheses are based on 10 or fewer galaxy redshifts and should be considered estimates.}
\begin{tabular}{@{\hspace{0mm}}l@{\hspace{3mm}}l@{\hspace{3mm}}c@{\hspace{3mm}}c@{\hspace{3mm}}c@{\hspace{3mm}}c@{\hspace{3mm}}c@{\hspace{3mm}}c@{\hspace{3mm}}c@{\hspace{3mm}}c@{\hspace{3mm}}c@{\hspace{0mm}}}\hline
&  &  & FSC/ &  &  &  & \multicolumn{2}{c}{$L_{\rm X}$} & &\\ 
Name & Other name & R.A.\ \& Dec (J2000) & BSC & $n_{\rm phot}$ & $f_{\rm X,RASS}$ & $z$ & RASS &CXO & $\sigma$ & $z$ ref\\ \hline
eMACS\,J0840.2+4421 & & 08 40 14.0  \,\,$+$44 21 53& F & 15 &1.13 & 0.6393 & 14.4 & & 1310 &(1)\\
RXJ0959.0+0255 & MaxBCG\,J149.94873+00.81880 &  09 59 02.8  \,\,$+$02 55 37 & F & 14 & 0.53 & 0.3494 & \,\,\,1.9 & & \,\,\,590 &(1)\\  
eMACS\,J1057.5+5759 & SL\,J1057.5+5759 & 10 57 35.0  \,\,$+$57 59 35 & B & 28 & 0.76 &0.5978 & \,\,\,8.5 & & \,\,\,860 &(1)\\
RXJ1411.3+5212 &  3C\,295, MACS\,J1411.3+5212 & 14 11 21.2  \,\,$+$52 12 50 & B & 51 & 1.07 & 0.4600 & \,\,\,6.7 & 10.1 & 1570 &(2)\\
eMACS\,J1419.2+5326  & RCS\,J141910+5326.2 & 14 19 15.2 \,\,$+$53 26 44 & F & 13 & 0.27 & 0.6384 & \,\,\,3.6 & \,\,\,3.2 & 1020 &(1) \\ 
RXJ1610.7+5406  & WHL\,J161040.5+540630 & 16 10 46.5  \,\,$+$54 06 55 & F & \,\,\,8 & 0.11 & \,\,\,0.3375$^a$ & \,\,\,0.4 & $<$0.5\,\, & (590) &(3)\\
RXJ1611.5+5417 &  WHL J161135.9+541634 & 16 11 34.7  \,\,$+$54 17 04 & F & 28 & 0.37 & 0.3381 & \,\,\,1.2 & \,\,\,2.3 & (810) & (3)\\   
RXJ1613.7+5542  &  WHL J161342.1+554155 & 16 13 42.3  \,\,$+$55 42 02 & F & 34 & 0.39 & 0.3512 & \,\,\,1.4 & {\it Cont.} & (590) &(1)\\
eMACS\,J1614.1+5404 &  & 16 14 06.4  \,\,$+$54 04 09 & F & \,\,\,7 & 0.11 & (0.65\,\,\,\,\,\,) & \,\,\,1.7 & & &\\ 
RXJ1614.2+5442 &  RX\,J1614.2+5442 & 16 14 15.1  \,\,$+$54 42 47 & B & 60 & 0.82 & 0.331\,\,\, & \,\,\,2.6 & {\it Cont.} & &(4)\\ 
eMACS\,J1616.7+5545  & SpARCS\,J161641+554513 & 16 16 43.9  \,\,$+$55 45 55 & F & 28 & 0.34 & 1.161\,\,\, &  16.2 & \,\,\,3.3 & (920) & (5) \\ \hline
\end{tabular}
\parbox{\textwidth}{{\it Note.} Redshift references: (1) this work, (2) Mann \& Ebeling (2012), (3) Trichas et al.\ (2010), (4) Edge et al.\ (2003), (5) Demarco et al.\ (2010); redshifts in parentheses are photometric estimates.  Point-source corrected X-ray luminosities are listed where suitable {\it Chandra}\/ data are available; we use the $L_{\rm X, CXO}$ column also to qualitatively mark two clusters as contaminated by AGN emission based on optical or X-ray evidence (see also Sections~\ref{sec:sample} and \ref{sec:cxo}). $^a$ BCG redshift.}
\end{table*}

\begin{figure*}
\hspace*{-2mm}
\includegraphics[width=0.44\textwidth]{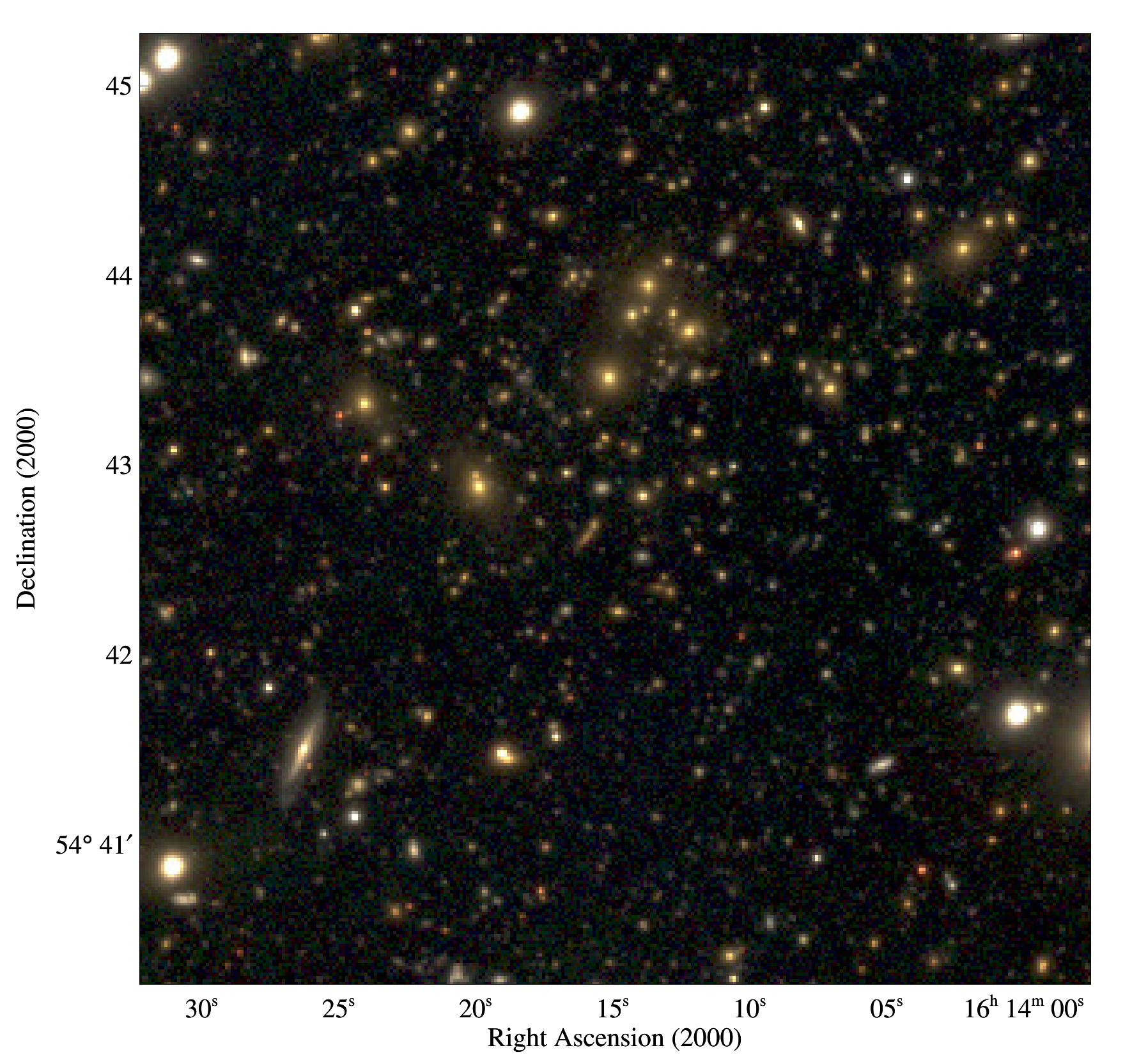}\hspace*{-3mm}
\includegraphics[width=0.44\textwidth]{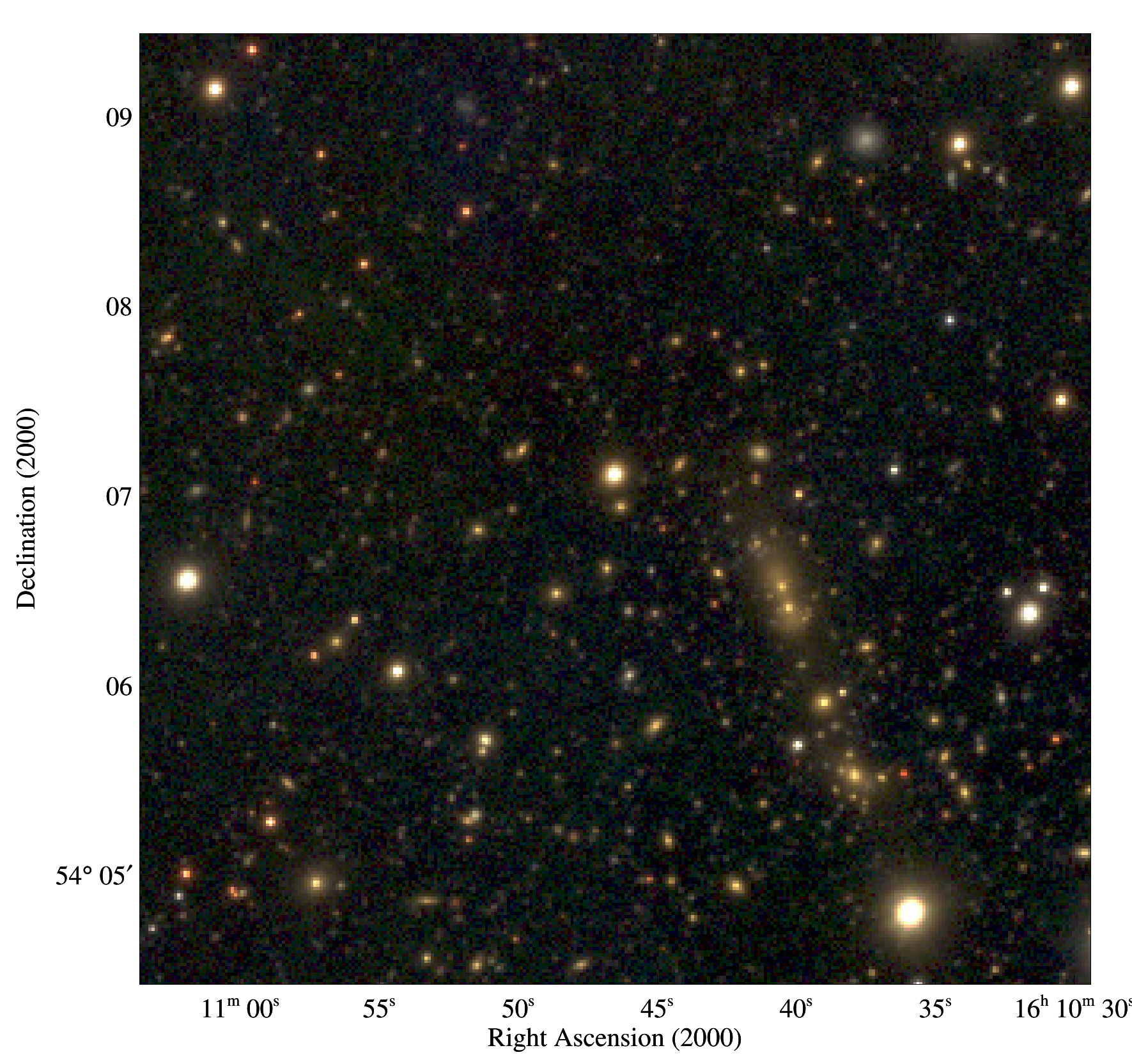}
\includegraphics[width=0.44\textwidth]{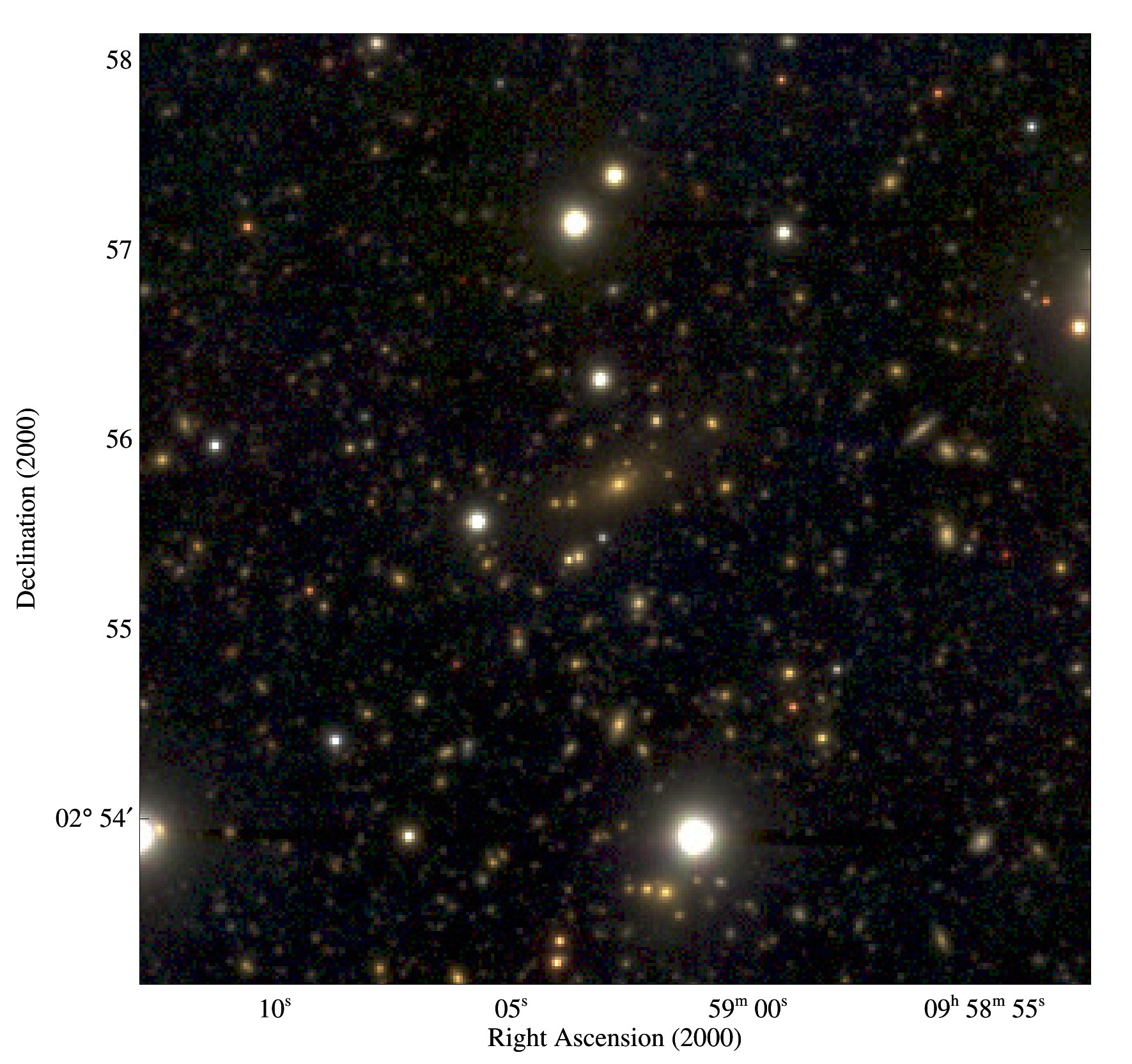}\hspace*{-3mm}
\includegraphics[width=0.44\textwidth]{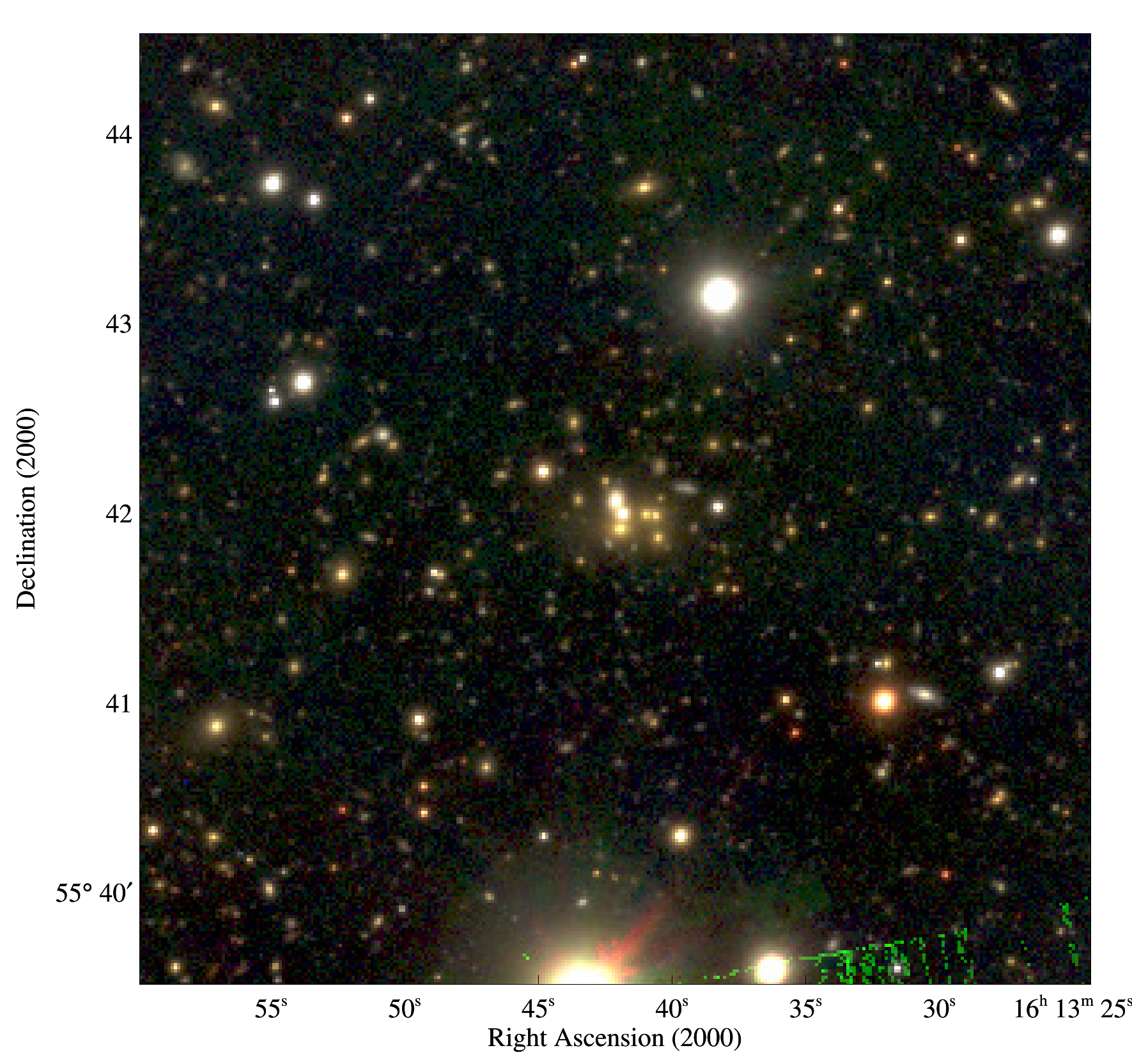}
\includegraphics[width=0.44\textwidth]{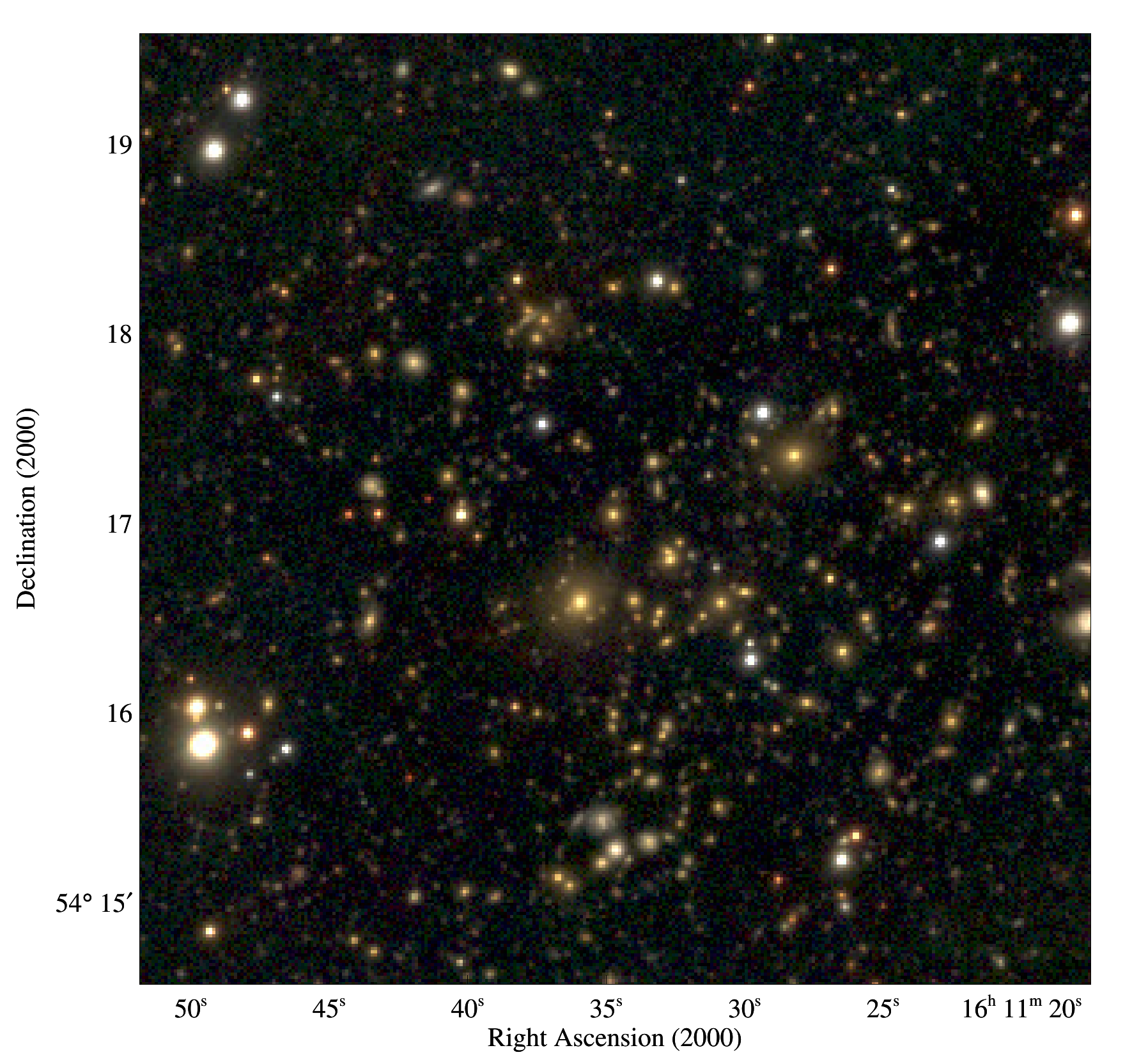}\hspace*{-3mm}
\includegraphics[width=0.44\textwidth]{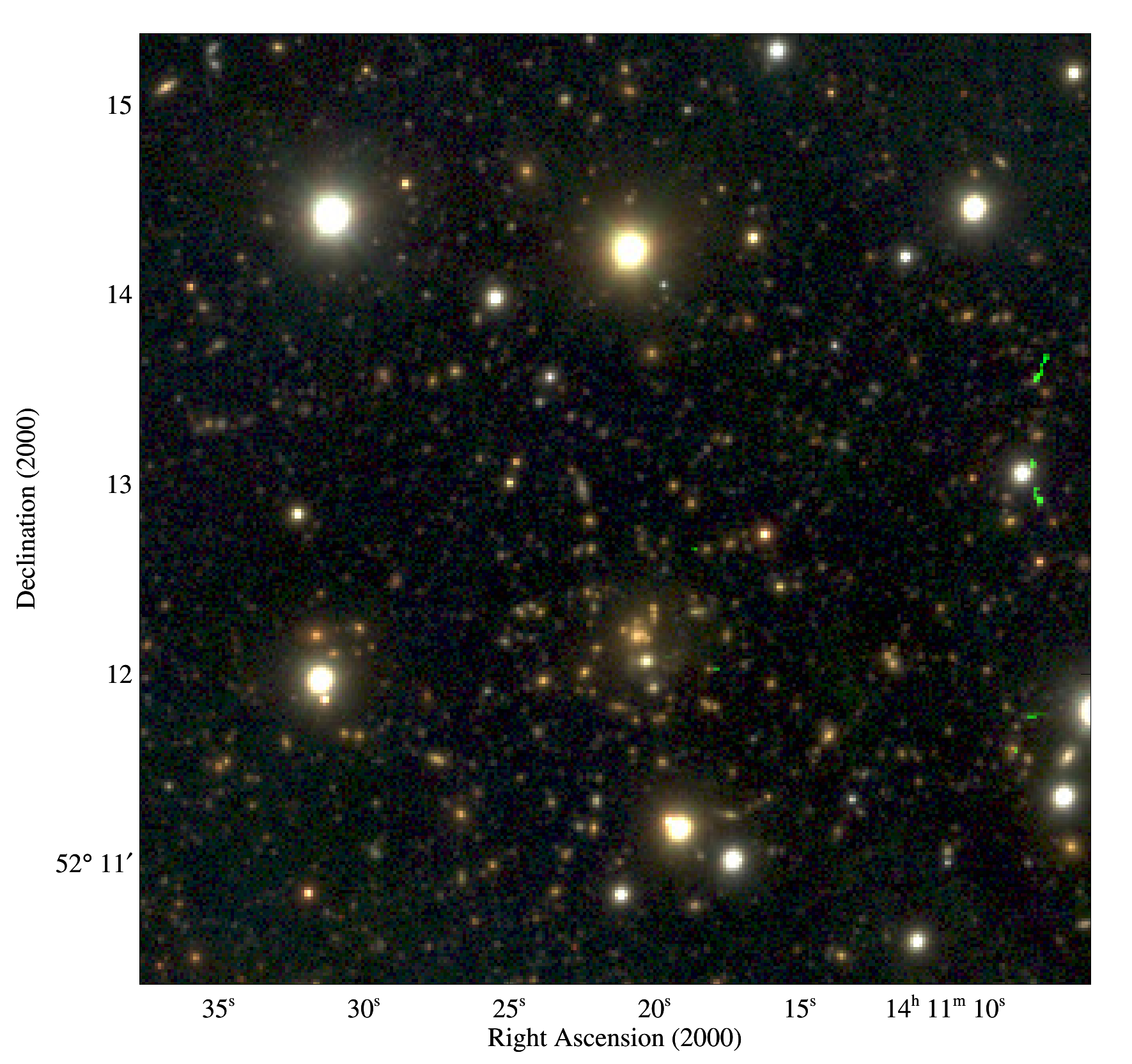}
\caption{PS1/MDS color images of a $5\times 5$ arcmin$^2$ region around the RASS sources identified as  clusters at $z>0.3$ in our eMACS pilot study. Clusters are shown in order of increasing redshift. The observed offsets of 30--60'' between the RASS source position and the location of the brightest cluster galaxy are fully consistent with the RASS point-spread function and with the typical positional uncertainties found in the course of the MACS project. From top to bottom and left to right: RXJ1614.2+5442 ($(gri)_{\rm P1}$, $z=0.33$), RXJ1610.7+5406 ($(gri)_{\rm P1}$, $z=0.34$), RXJ0959.0+0255 ($(gri)_{\rm P1}$, $z=0.35$), RXJ1613.7+5542 ($(gri)_{\rm P1}$, $z=0.35$), RXJ1611.5+5417 ($(gri)_{\rm P1}$, $z\sim0.38$), and RXJ1411.3+5212 ($(gri)_{\rm P1}$, $z=0.46$). \label{fig:mdim}}
\end{figure*}

\begin{figure*}
\hspace*{-2mm}
\includegraphics[width=0.44\textwidth]{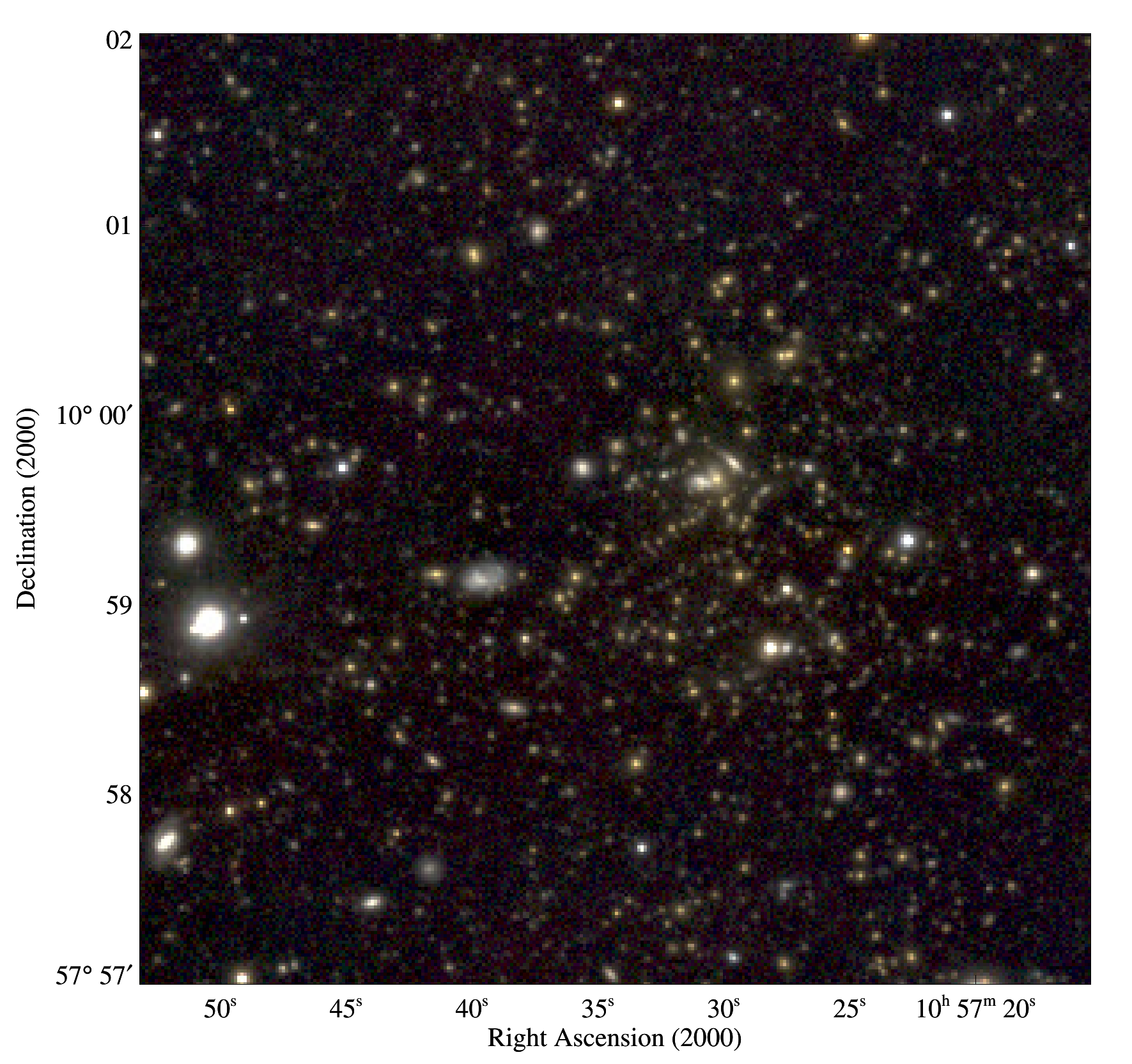}\hspace*{-3mm}
\includegraphics[width=0.44\textwidth]{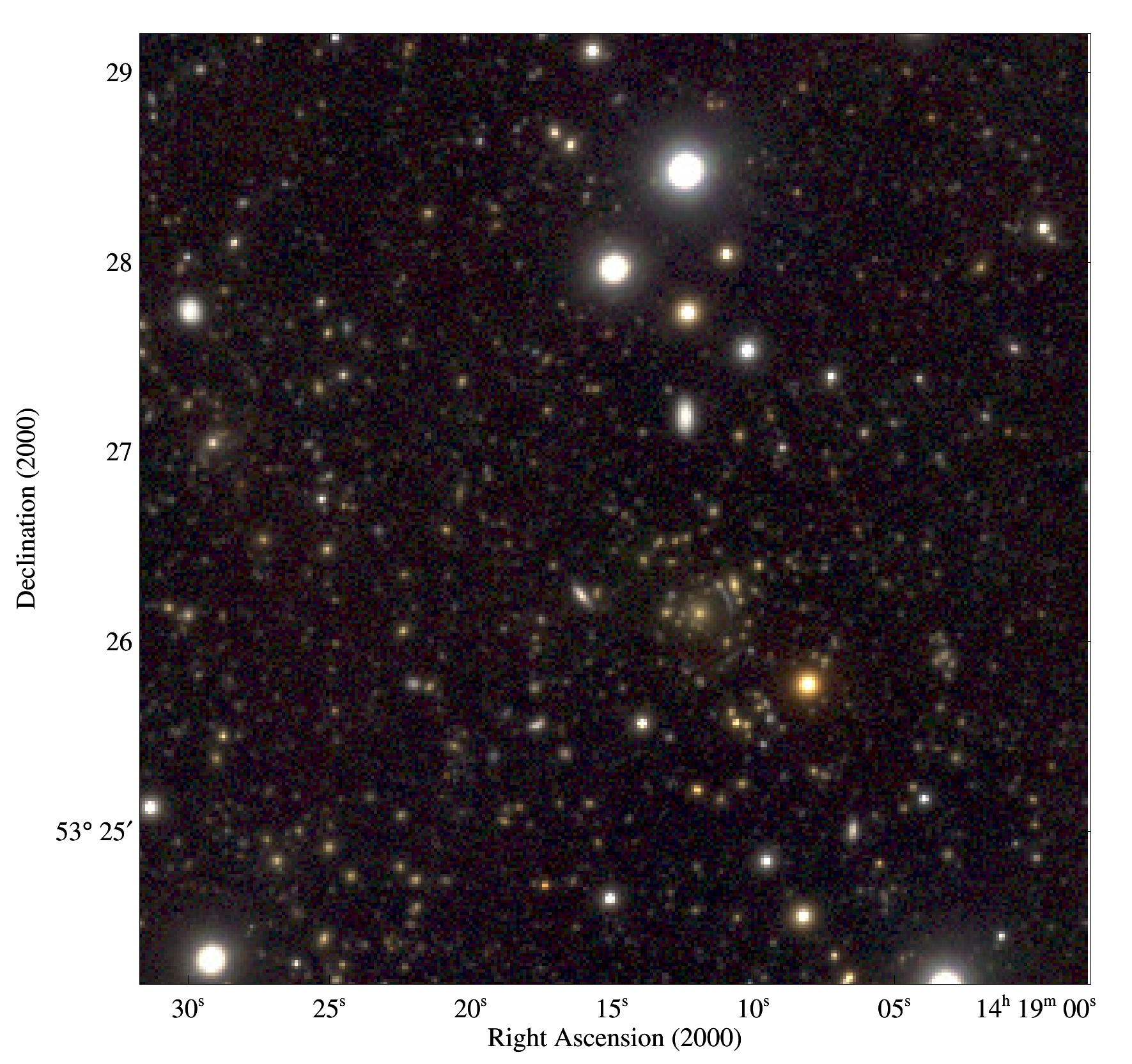}
\includegraphics[width=0.44\textwidth]{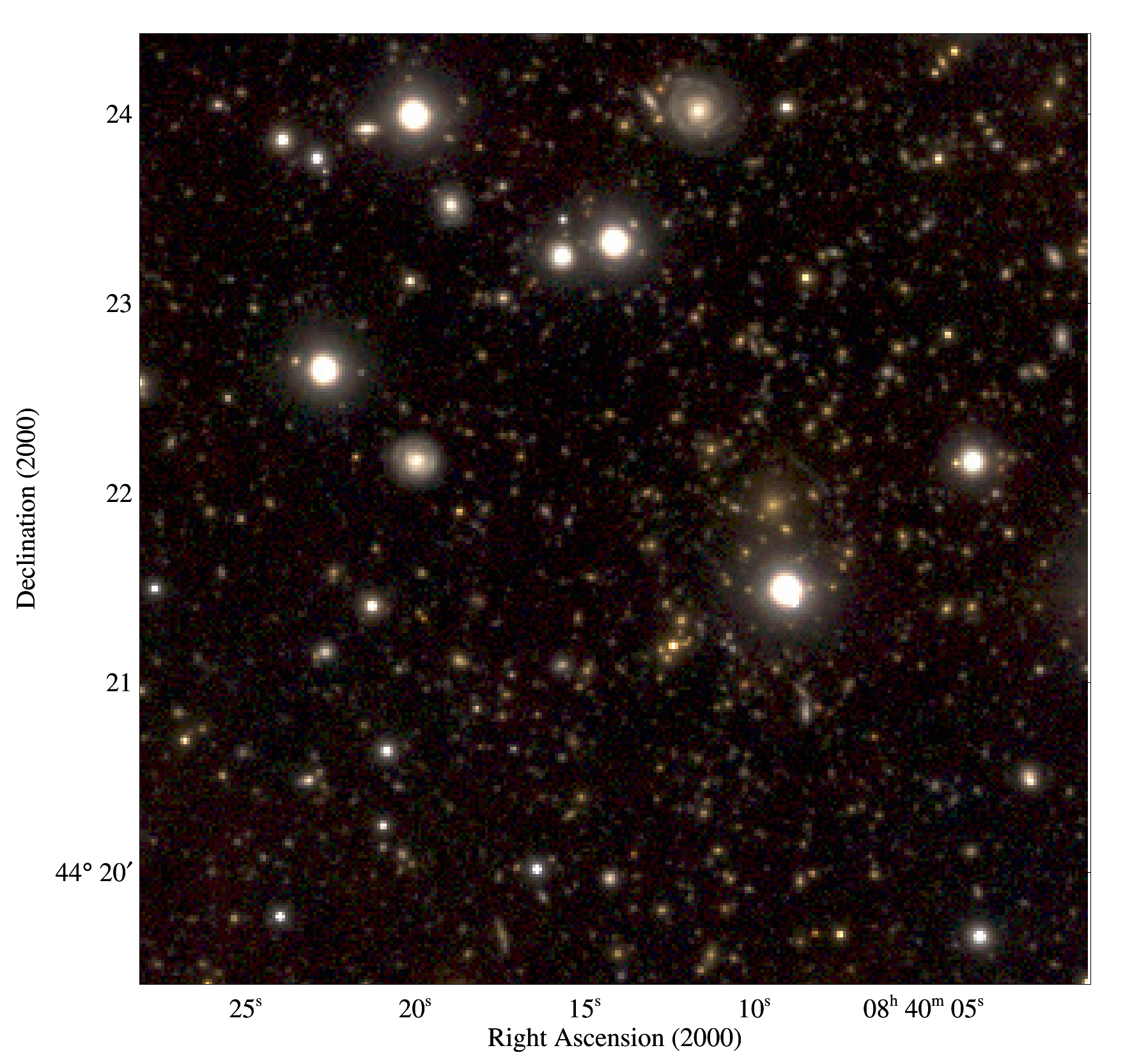}\hspace*{-3mm}
\includegraphics[width=0.44\textwidth]{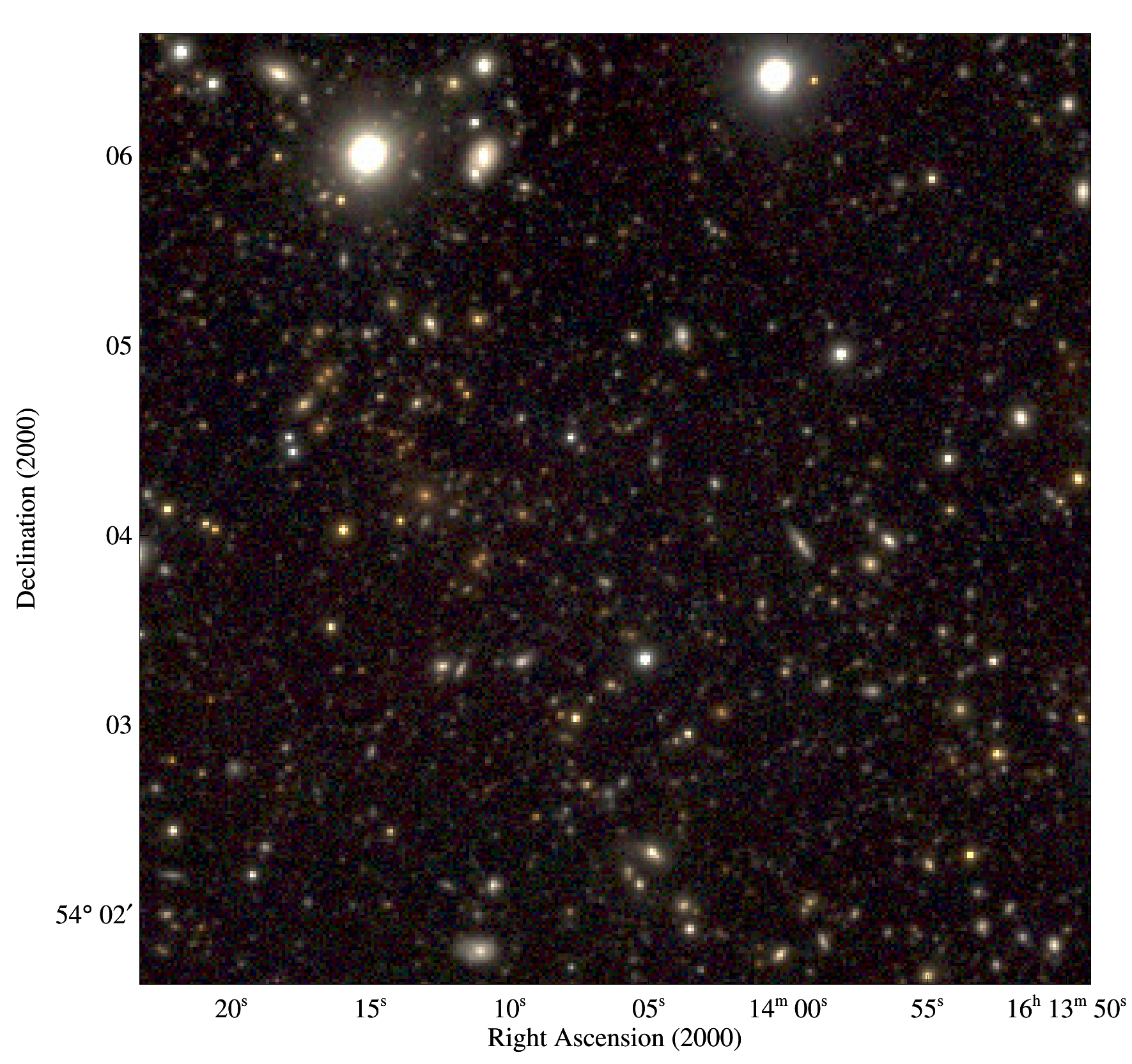}
\includegraphics[width=0.44\textwidth]{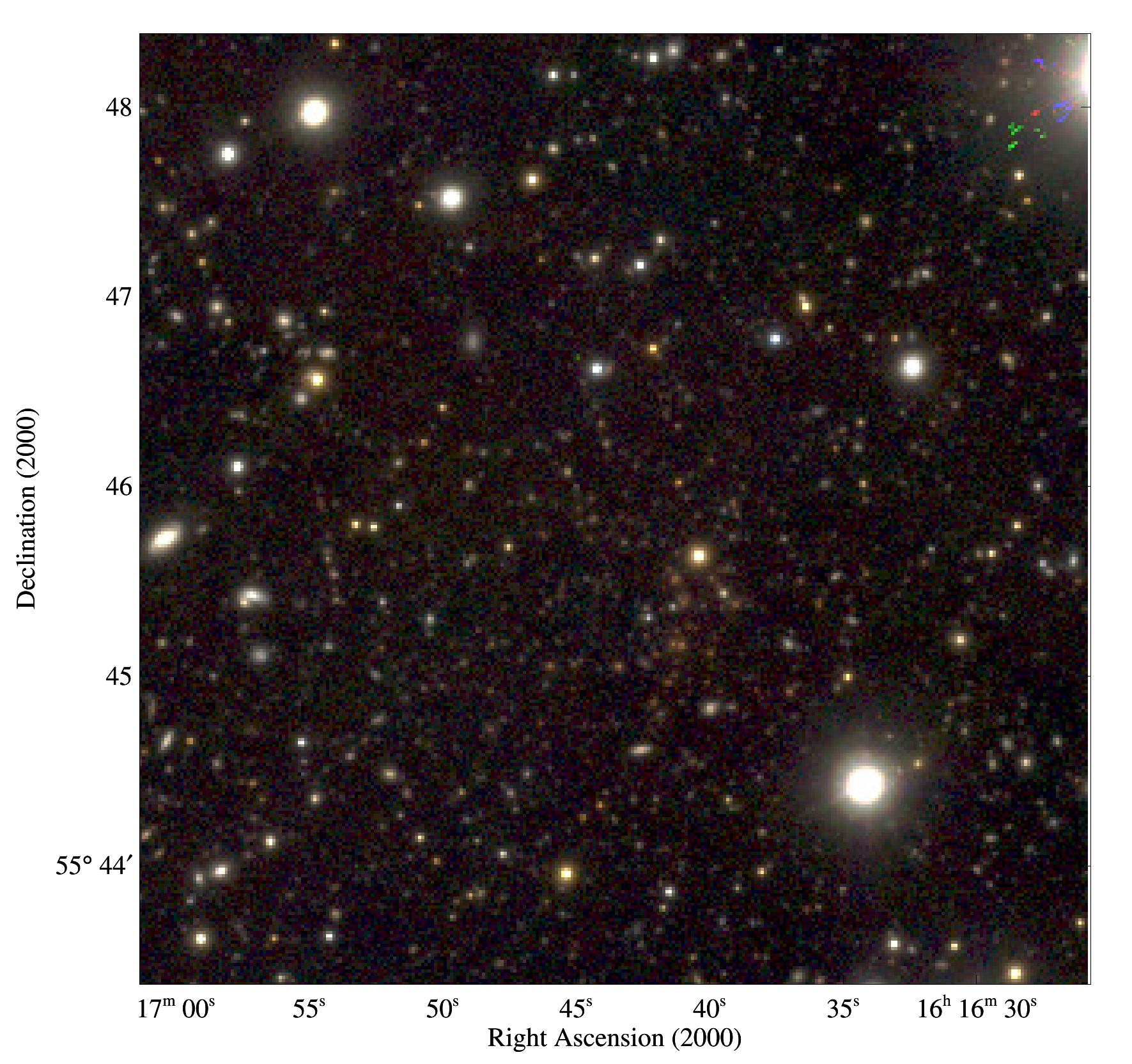}\\
\contcaption{From top to bottom and left to right:  eMACS\,J1057.5+5759 ($(riz)_{\rm P1}$, $z=0.60$), eMACS\,J1419.2+5326  ($(riz)_{\rm P1}$, $z=0.64$), eMACS\,J0840.2+4421 ($(riz)_{\rm P1}$, $z=0.64$), eMACS\,J1614.1+5404 ($(riz)_{\rm P1}$, $z\sim0.65$), and eMACS\,J1616.7+5545 ($(riz)_{\rm P1}$, $z=1.16$).}
\end{figure*}

\begin{figure*}
\hspace*{-2mm}
\includegraphics[width=0.44\textwidth]{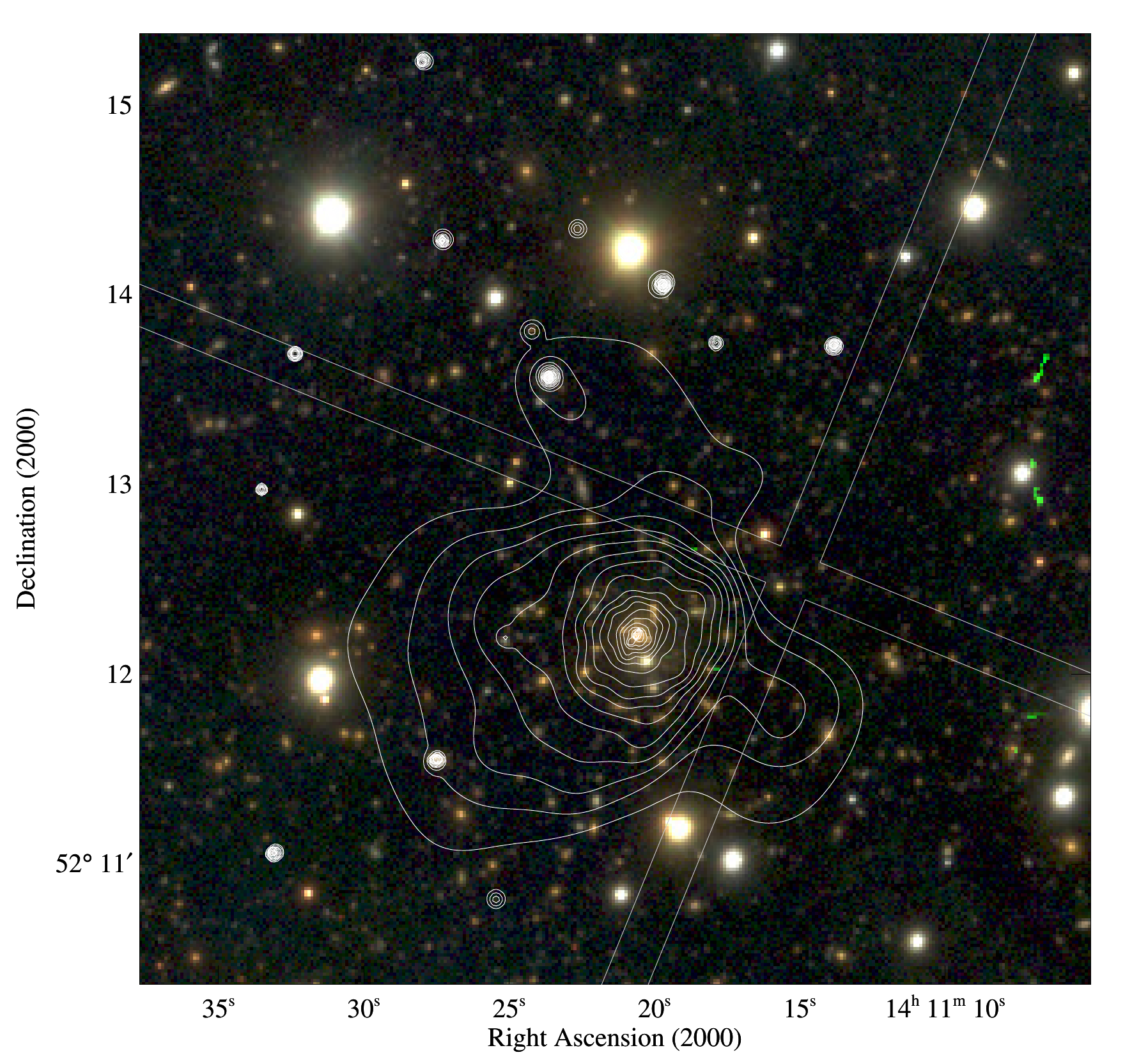}\hspace*{-3mm}
\includegraphics[width=0.44\textwidth]{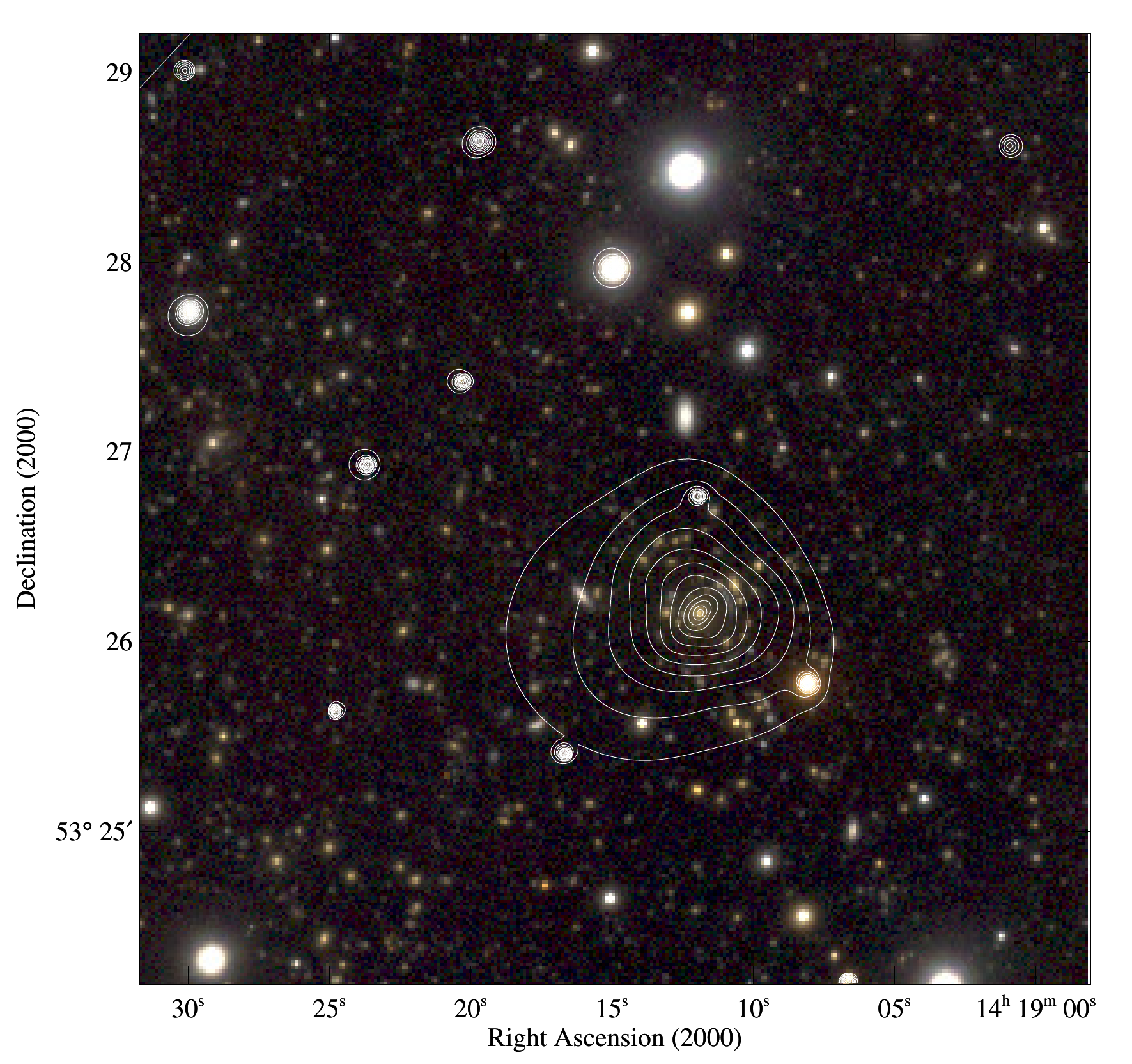}
\includegraphics[width=0.44\textwidth]{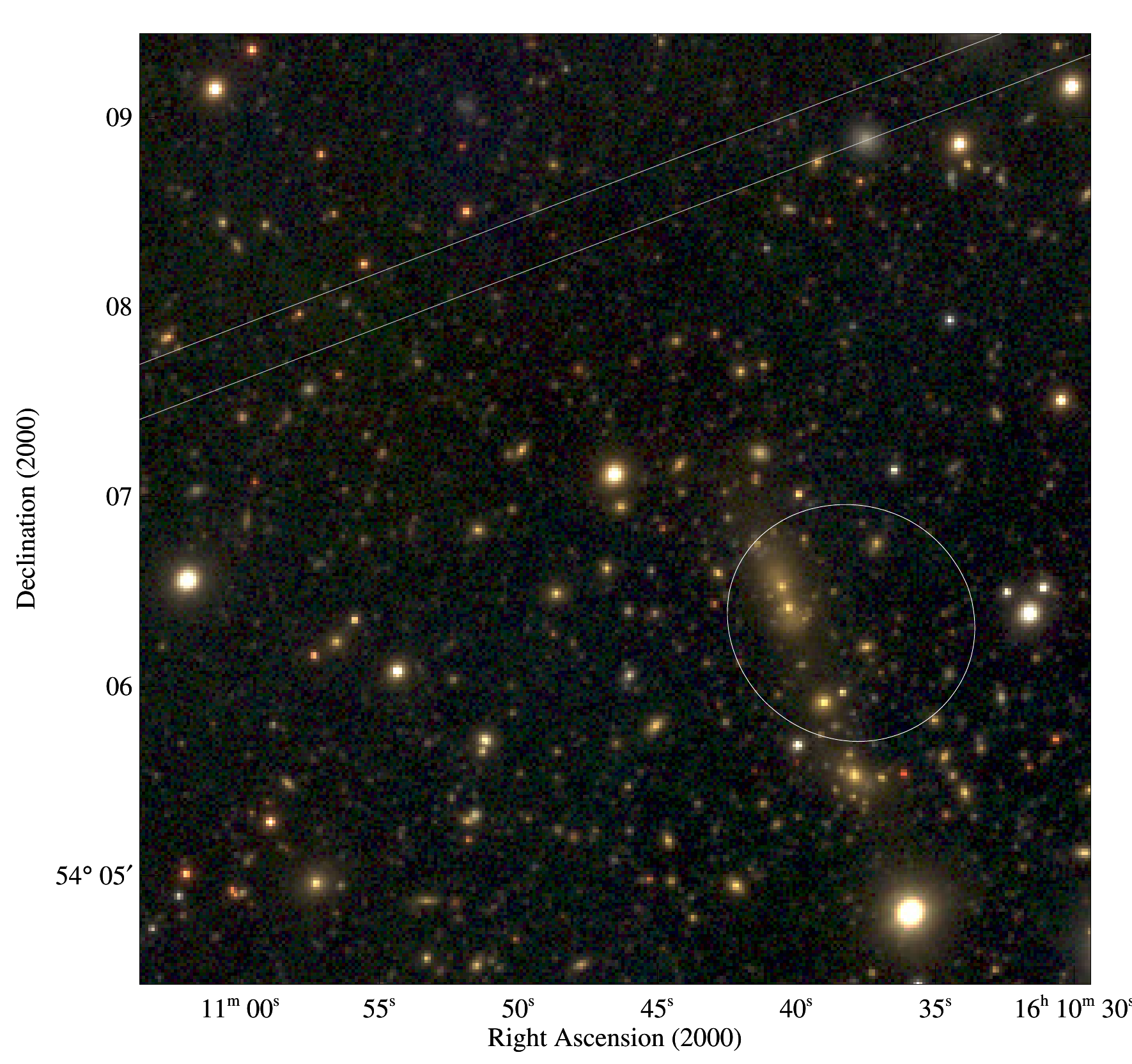}\hspace*{-3mm}
\includegraphics[width=0.44\textwidth]{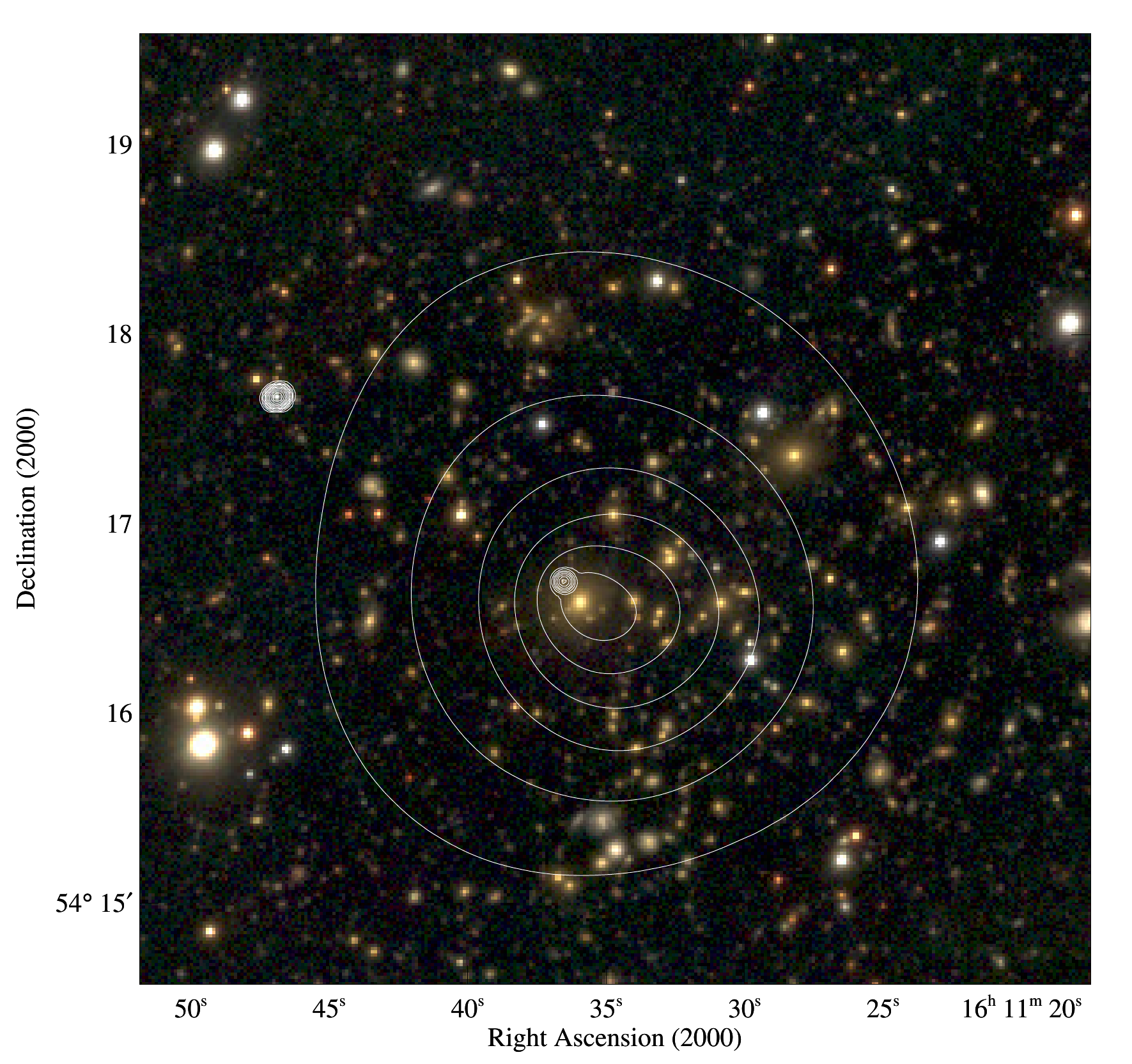}
\includegraphics[width=0.44\textwidth]{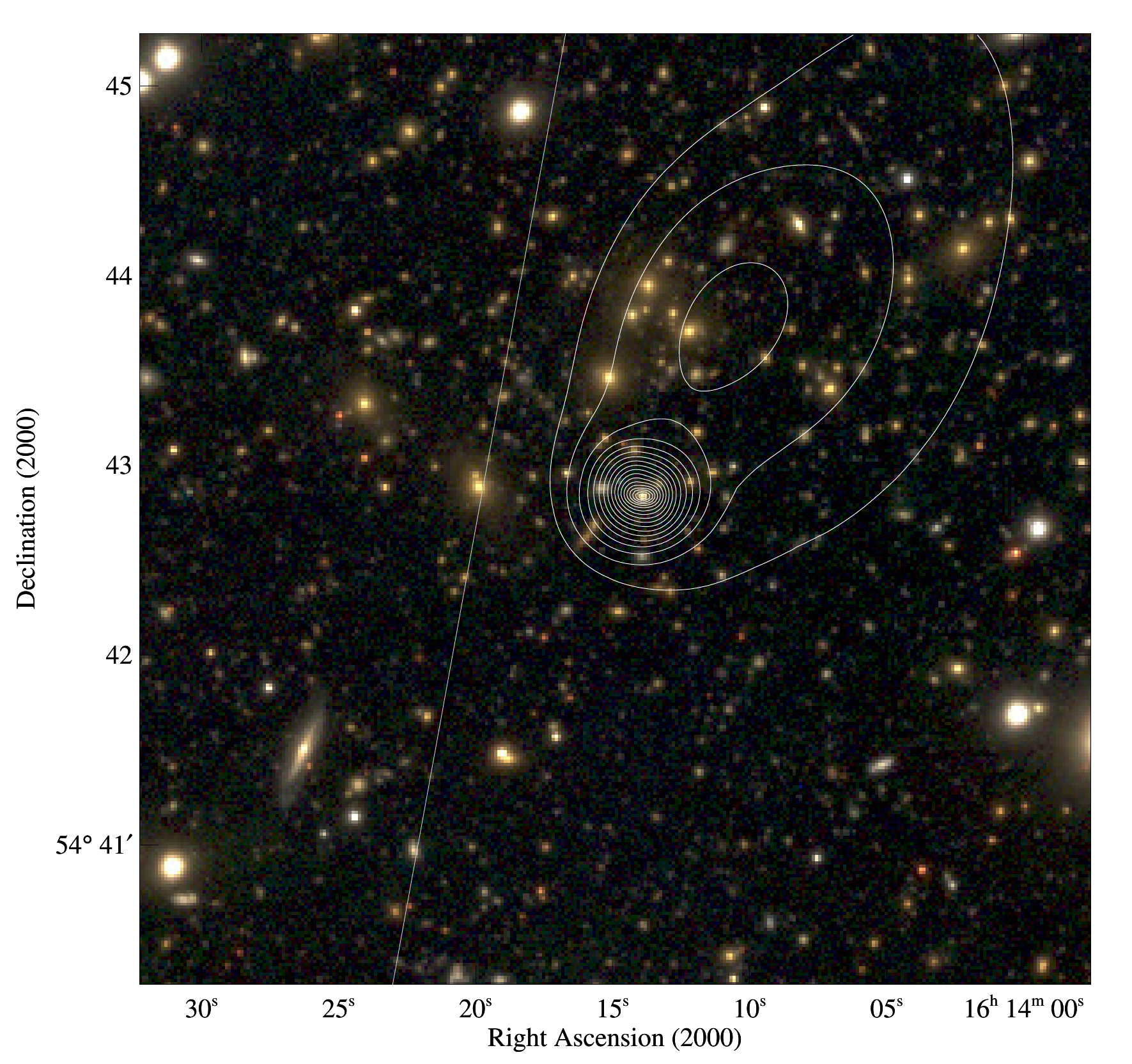}\hspace*{-3mm}
\includegraphics[width=0.44\textwidth]{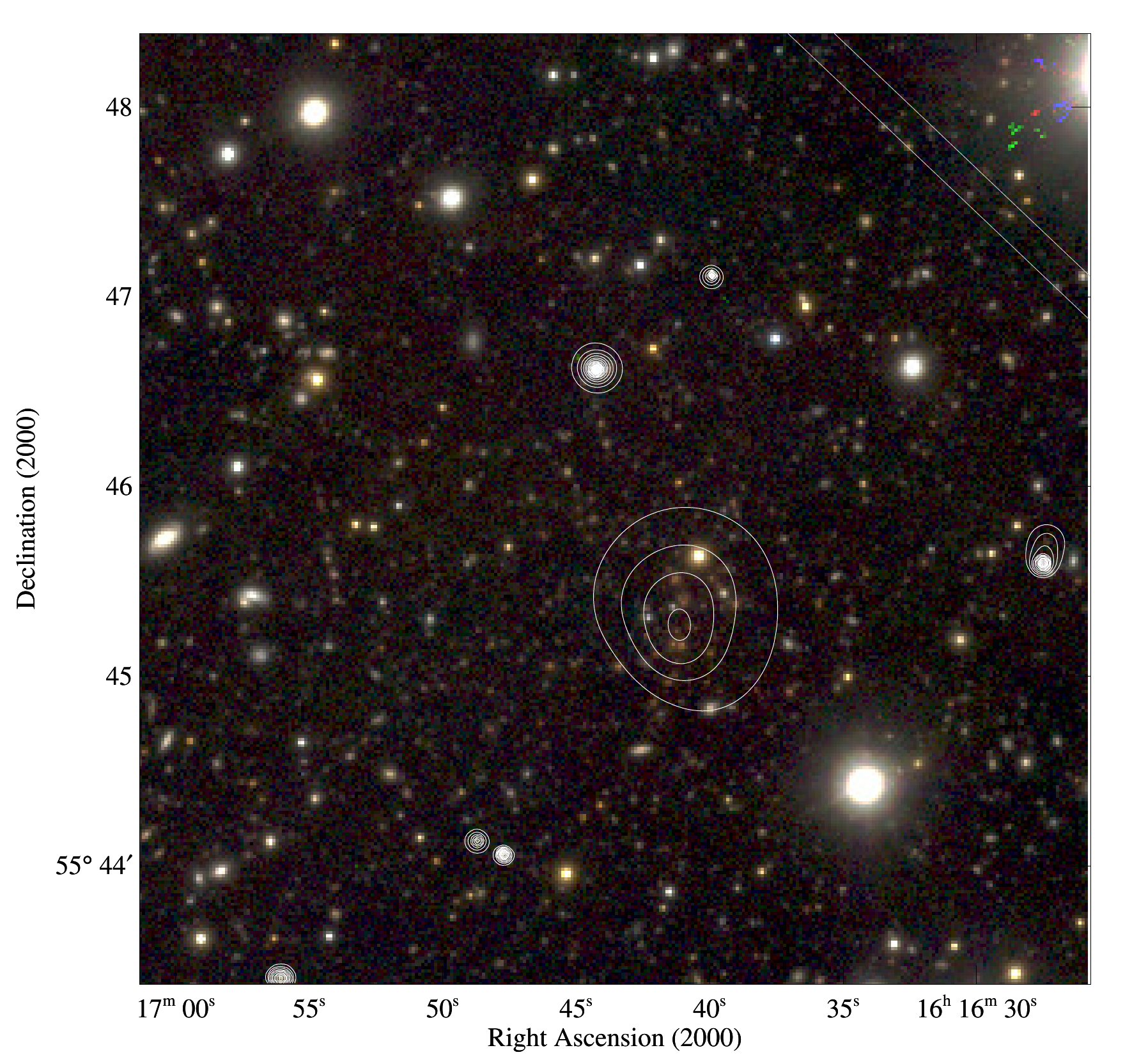}
\caption{As Fig.~\ref{fig:mdim}, but only for the six clusters with targeted or serendipitous {\it Chandra}\/ data. Overlaid are surface-brightness contours of the adaptively smoothed X-ray emission from (top to bottom and left to right) RXJ1411.3+5212, eMACS\,J1419.2+5326,  RXJ1610.7+5406, RXJ1611.5+5417, RXJ1614.2+5442, and  eMACS\,J1616.7+5545. Contours are spaced logarithmically, with the lowest one being 50\% above the background level. Only part of the diffuse cluster emission is captured for RXJ1614.2+5442 which was serendipitously observed at the very edge of the Advanced CCD Imaging Spectrometer (ACIS-I) field of view (chip edges are marked by straight lines). Similarly, extended emission from RXJ1411.3+5212 (3C295) is affected by the cluster core being too close to the ACIS-I chip gaps. Note the presence of X-ray point sources in all fields. \label{fig:xim}}
\end{figure*}

\section{X-ray follow-up observations}\label{sec:cxo}

As discussed in Section~\ref{sec:xps}, contamination from X-ray point sources affects all X-ray cluster surveys, although its impact is (statistically) less severe for collections of extremely massive clusters. For individual RASS-discovered systems, however, the presence and degree of point-source contamination can be quantified only by X-ray follow-up observations. The facility of choice for this purpose is the {\it Chandra}\/ Observatory, whose 0.5\arcsec on-axis resolution allows the unambiguous identification (and removal) of point sources even in relatively short observations.

For six of the 11 clusters detected in our eMACS pilot project, {\it Chandra}\/ data are already available. One of them is RXJ1411.3+5212, the well studied cluster around the powerful radio source 3C\,295 (Allen et al.\ 2001). RXJ1610.7+5406, RXJ1611.5+5417, and RXJ1614.2+5442 were serendipitously observed during a shallow {\it Chandra}\/ survey of AGN in the European Large-Area {\it ISO}\/ Survey North 1 (ELAIS-N1) field (MD08).  The remaining two clusters, eMACS\,J1419.2+5326 and eMACS\,J1616.7+5545, are previously identified, optically selected clusters for which high-resolution {\it Chandra}\/ observations were awarded to the original discoverers. X-ray/optical overlays for all six clusters are shown in Fig.~\ref{fig:xim}. 

Of the five clusters observed with {\it Chandra}, two are found to feature point-source corrected luminosities  that fall significantly below the RASS estimates (see Table~\ref{tab:mdclusters}). For a third one (RXJ1614.2+5442) the coverage of the cluster by the existing (serendipitous) {\it Chandra}\/ data is insufficient to allow a reliable flux measurement; the data show clearly though that the RASS flux is dominated by a bright point source (Fig.~\ref{fig:xim}).  Of the two clusters at $z{>}0.5$ that have archival {\it Chandra}\/ data, the most distant one (eMACS\,J1616.7+5545 at $z{=}1.161$) is found to be a blend of a bright, unrelated QSO and a  moderately X-ray luminous cluster. {\it Chandra}\/ observations also confirm eMACS\,J1419.2+5326 at $z{=}0.638$ as intrinsically X-ray luminous; however, it too is significantly contaminated by several X-ray point sources. Of the three point sources closest to the core of eMACS\,J1419.2+5326, one is a star, and the other two are background QSOs (see Fig.~\ref{fig:xim} and Table~\ref{tab:redshifts}). Again, the contaminating point sources are thus projected onto the cluster, as assumed in the probabilistic argument put forward in Section~\ref{sec:xps}. 

Figure~\ref{fig:mdim} shows offsets of, on average, almost 40\arcsec\ (corresponding to over 160 kpc at $z{>}0.3$ and over 220 kpc at $z{>}0.5$) between the position of a RASS X-ray source and the location of the BCG or, more generally, the cluster core. The reasons for such misalignments, which are much larger than those observed for true X-ray point sources (Fig.~\ref{fig:xpsoff}), are threefold. For one, the intrinsically extended emission from clusters, in particular of unrelaxed systems, causes larger positional uncertainties, even if, at the resolution of the RASS, the extent of the emission is usually not resolved at $z{>}0.3$. The two other causes of the X-ray / optical offsets apparent in Fig.~\ref{fig:mdim} are illustrated by Fig.~\ref{fig:xim}. In most cases (RXJ1411.3+5212, eMACS\,J1419.2+5326, RXJ1614.2+5442, and  eMACS\,J1616.7+5545), blends of X-ray point sources and the cluster emission are responsible for the observed X-ray / optical misalignment. For very faint RASS sources, however, similarly large offsets can simply be the result of exceedingly poor photon statistics in the RASS (RXJ1610.7+5406, eight net RASS photons). 

\section{Cluster velocity dispersions}\label{sec:veldisp}

As pointed out in the previous section, a significant distance between a RASS source from the BCG of its presumed cluster counterpart may be indicative of the presence of contaminating X-ray point sources. Point-source contamination can, however, be significant also for clusters that show only small X-ray / optical offsets.

Cluster velocity dispersions could, in principle, be used as an alternative way to flag systems that are unlikely to be as X-ray luminous (or massive) as suggested by their RASS flux since, for virialized systems, the dispersion of the radial velocities of cluster galaxies is well correlated with cluster mass (Carlberg et al.\ 1996). However, most clusters, in particular at moderate to high redshift, are still growing and often observed before, during, or after significant merger events (e.g., Mann \& Ebeling 2012). For such systems, the observed velocity dispersion can be boosted well beyond the virial value. In addition, observational biases (viewing angle, unrepresentative sampling) can lead to either over- or underestimates of the virial mass. As a result, cluster velocity dispersions should be used with great caution when mass estimates are sought for individual clusters.

We list in Table~\ref{tab:mdclusters}  velocity dispersions, $\sigma$, based on galaxy redshifts obtained either from the literature or from our own measurements, as tabulated in Table~\ref{tab:redshifts}. For systems with fewer than 10 redshifts, the listed values should be considered rough estimates. We find velocity dispersions ranging from approximately 600 km $s^{-1}$ for several moderately X-ray luminous clusters at $z{\sim} 0.35$ to over 1500 km $s^{-1}$ for the very X-ray luminous 3C\,295 cluster (MACS\,J1411.3+5212). 

Unsurprisingly, we find the $\sigma$ values listed in Table~\ref{tab:mdclusters} to be poorly correlated with  cluster X-ray luminosity. Indeed, the 12 most distant MACS clusters (all of them confirmed to be massive systems at $z{>}0.5$) currently have measured velocity dispersions between 750 and over 1,600 km s$^{-1}$. This wide range of velocity dispersions encompasses all of the $\sigma$ values listed in Table~\ref{tab:mdclusters} for cluster detections in the same redshift range from our eMACS pilot study. We conclude that velocity dispersions may allow the elimination of some low-mass systems from a sample of eMACS candidate clusters; serious disadvantages of velocity dispersions are that they are expensive to obtain and noisy as predictors of X-ray luminosity (or mass).

\section{Summary and Discussion} \label{sec:discussion}
  
Our eMACS pilot study conducted within the 71 deg$^2$ of the PS1 MDS confirms the premise of the eMACS project that even the faintest sources from the RASS BSC and FSC catalogues can be used to successfully identify distant, X-ray luminous clusters. X-ray follow-up observations with {\it Chandra}, however, also confirm that contamination from X-ray point sources is common and often severe at such low flux levels. This is a particular concern at the highest redshifts probed by eMACS, where clusters of low to moderate mass can masquerade as exceptionally X-ray luminous by virtue of being blended with nearby, X-ray bright point sources\footnote{SpARCS\,J161641+554513 at $z{=}1.161$ provides a striking example of this effect. }. While focusing on the optically richest candidates helps to prevent misidentifications, the ubiquity of X-ray luminous QSOs makes high-resolution follow-up with {\it Chandra}\/ a necessity if accurate cluster luminosities are to be established.

Since point-source contamination and blends in the RASS can only  boost but never lower the RASS flux of  eMACS cluster candidates, we interpret $L_{\rm X,RASS}+dL_{\rm X,RASS}$ as a {\it de facto} upper limit to the true cluster flux and adopt $L_{\rm X,RASS}+dL_{\rm X,RASS}> 5\times 10^{44}$ erg s$^{-1}$ as a luminosity requirement that must be met by any eMACS cluster candidate selected from the FSC or BSC at a measured or estimated redshift of $z{>}0.5$. Of the 11 clusters from our pilot-study sample, four meet this requirement. It is clear though from simple scaling arguments (four eMACS candidates in 71 deg$^2$ imply a yield of over 1,000 such clusters within the full eMACS solid angle of over 20,000 deg$^2$) that the majority of the clusters thus selected can not conceivably be as X-ray luminous as suggested by their RASS fluxes. Consistent with this expectation, {\it Chandra}\/ observations of two of the four eMACS candidate clusters identified in our pilot study\footnote{We note that the {\it Chandra}\/ observations for both of these clusters were awarded as the result of these systems' independent discovery in optical cluster surveys.} indeed found bright X-ray point sources in the immediate vicinity of either cluster, resulting in point-source corrected luminosities that fall well short of the eMACS target of $L_{\rm X}\ga 1\times 10^{45}$ erg s$^{-1}$. 

To reduce the prohibitively (and erroneously) large number of seemingly eMACS-like clusters created by the superposition of unrelated X-ray point sources, we investigate optical cluster properties in search of an additional selection criterion; our aim being to identify the clusters among our candidates that have the highest probability of truly being as X-ray luminous as implied by their RASS fluxes. The data collected in our pilot study and presented in Figs.~\ref{fig:mdim} and \ref{fig:xim}, as well as in Table~\ref{tab:redshifts}, prove instructive in this regard. Adopting a threshold  of $\sigma{>}750$ km s$^{-1}$ for the cluster velocity dispersion eliminates clusters of low to moderate mass, but cannot discriminate between genuinely massive clusters and poorer systems whose velocity dispersion is boosted by ongoing mergers or infall along our line of sight\footnote{An example of such orientation bias for an evolving cluster is RCS\,J141910+5326.2 (eMACS\,J1419.2+5326) at $z=0.64$ which features a velocity dispersion of over 1,000 km s$^{-1}$ but falls well short of eMACS X-ray luminosity requirements.}. A complementary criterion is provided by the offset between the RASS source position and the position of the BCG. For all but the least significant RASS detections, offsets exceeding 300 kpc at the cluster redshift point to the likely presence of contaminating X-ray point sources. Finally, a third screening criterion can be obtained by simple visual inspection. Imaging data obtained for MACS (e.g., Ebeling et al.\ 2007, 2010), or of the PS1/MD images of the five eMACS candidate clusters at $z{>}0.5$ shown in Fig.~\ref{fig:mdim}, show that, at the extreme end of the cluster mass function probed by MACS and eMACS, optical richness becomes a powerful discriminator between truly X-ray luminous clusters and intrinsically poorer systems. RCS\,J141910+5326.2 and eMACS\,J0840.2+4421, both at $z{=}0.64$, differ dramatically in this regard (Fig.~\ref{fig:mdim}),  the former appearing highly compact and optically poor compared to the latter. However, just like velocity dispersion, optical richness too is an unreliable indicator of mass, in the sense that it fails to select highly evolved cool-core clusters (note the unimpressive appearance of the cluster around 3C\,295 in Fig.~\ref{fig:mdim}). A more robust predictor of cluster mass than optical richness, traditionally understood to be the number of cluster galaxies above a certain magnitude threshold, is the total stellar mass in cluster galaxies (Andreon 2012). However, the respective correlation is presently not well calibrated at high cluster masses.

We conclude that the only reliable way of eliminating impostors from the list of eMACS candidate clusters at $z{>}0.5$ is a {\it Chandra}\/ snapshot observation. Since performing {\it Chandra}\/ observations of all candidates would be prohibitively expensive, the most promising targets need to be selected using velocity dispersions, X-ray / optical offsets, and optical appearance. Applying these criteria retroactively to RCS\,J141910+5326.2 and SpARCS\,J161641+554513 (the two eMACS candidates already observed with {\it Chanda}) we find both of them to appear at best moderately rich in the optical and to exhibit X-ray / optical offsets that strongly suggest contamination from X-ray point sources -- as confirmed by {\it Chandra}. 

By contrast, eMACS\,J1057.5+5759 appears as optically rich as the most massive MACS clusters and features only a small X-ray / optical offset. eMACS\,J0840.2+4421 exhibits an optical morphology suggestive of a very massive, relaxed cluster and a commensurately high velocity dispersion of over 1,300 km s$^{-1}$. The sizeable offset of 340 kpc between its BCG and  the RASS X-ray position, however, suggests the presence of a contaminating X-ray point source to the east of the cluster.  

Although the solid angle covered by our pilot study is too small to allow a meaningful extrapolation to the over 20,000 deg$^2$ surveyed by eMACS proper,  the discovery of eMACS\,J0840.2+4421 and eMACS\,J1057.5+5759 bode well for the eMACS sample being compiled from RASS and PS1 $3\pi$ data. Establishing robust, point-source corrected X-ray luminosities with {\it Chandra}\/ for both of these systems would provide an important first test of the eMACS survey strategy.  

\section*{Acknowledgements}

HE gratefully acknowledges financial support from NASA/ADP grant NNX11AB04G. We thank Matthew Zargursky for developing code to merge and display PS1/MD imaging data during the early phase of this project. The authors wish to recognize and acknowledge the very significant cultural role and reverence that the summit of Mauna Kea has always had within the indigenous Hawaiian community.  We are most fortunate to have the opportunity to conduct observations from this mountain. 

The Pan-STARRS1 Surveys (PS1) have been made possible through contributions of the Institute for Astronomy, the University of Hawaii, the Pan-STARRS Project Office, the Max-Planck Society and its participating institutes, the Max Planck Institute for Astronomy, Heidelberg and the Max Planck Institute for Extraterrestrial Physics, Garching, The Johns Hopkins University, Durham University, the University of Edinburgh, Queen's University Belfast, the Harvard-Smithsonian Center for Astrophysics, the Las Cumbres Observatory Global Telescope Network Incorporated, the National Central University of Taiwan, the Space Telescope Science Institute, and the National Aeronautics and Space Administration under Grant No. NNX08AR22G issued through the Planetary Science Division of the NASA Science Mission Directorate and the University of Maryland.

\appendix
\section{Spectroscopic follow-up observations}\label{sec:appendix}

Five clusters from the sample listed in Table~\ref{tab:mdclusters}  that lacked secure spectroscopic redshifts prior to our project were targeted in dedicated follow-up observations from Mauna Kea. Spectra of galaxies presumed to be cluster members based on their colour in PS1 images, as well as of objects that might be the source of contaminating point-like X-ray emission, were obtained with the DEIMOS spectrograph (Faber et al.\ 2003) on the Keck-2 10m telescope. We used the 600 l/mm grism and the GG455 blocking filter to collect low-resolution spectra from 4500\AA\ to 9000\AA. Exposure times ranged from $3\times 600$s for clusters at $z\sim 0.3$ to $4\times 1800$ for our most distant targets at $z\sim 0.6$. All cluster redshifts thus obtained are listed in Table~\ref{tab:mdclusters}; the individual redshifts measured by us in the four cluster fields are tabulated in Table~\ref{tab:redshifts}. Also listed in Table~\ref{tab:redshifts} are the resulting velocity dispersions. 

\begin{table*}
\caption{\label{tab:redshifts} Galaxy redshifts obtained with Keck-II/DEIMOS during our eMACS pilot study.  }
\begin{tabular}{lcclcc}
Galaxy  & R.A.\ \& Dec (J2000) & $z$ & Galaxy  & R.A.\ \& Dec (J2000) & $z$ \\ \hline
eMACSJ0840.2+4421-g01   & 08 40 09.35 \,\,$+$44 21 54.1 &  0.6384 & eMACSJ0840.2+4421-g17   & 08 40 12.66 \,\,$+$44 23 55.7 &  0.6317 \\
eMAC,J0840.2+4421-g02   & 08 40 11.17 \,\,$+$44 22 11.6 &  0.6518 & eMACSJ0840.2+4421-g18   & 08 40 01.56 \,\,$+$44 24 35.9 &  0.6416 \\
eMACSJ0840.2+4421-g03   & 08 40 12.08 \,\,$+$44 21 17.9 &  0.6351 & eMACSJ0840.2+4421-g19   & 08 39 59.04 \,\,$+$44 23 39.1 &  0.6416 \\
eMACSJ0840.2+4421-g04   & 08 40 05.67 \,\,$+$44 22 43.2 &  0.6476 & eMACSJ0840.2+4421-g20   & 08 39 58.40 \,\,$+$44 22 55.3 &  0.6250 \\
eMACSJ0840.2+4421-g05   & 08 40 12.50 \,\,$+$44 21 06.1 &  0.6303 & eMACSJ0840.2+4421-g21   & 08 40 03.63 \,\,$+$44 20 04.5 &  0.6388 \\
eMACSJ0840.2+4421-g06   & 08 40 18.63 \,\,$+$44 21 05.8 &  0.6406 & eMACSJ0840.2+4421-g22   & 08 39 56.54 \,\,$+$44 21 04.1 &  0.5677 \\
eMACSJ0840.2+4421-g07   & 08 40 10.31 \,\,$+$44 22 46.0 &  0.6430 & eMACSJ0840.2+4421-g23   & 08 39 56.91 \,\,$+$44 22 02.6 &  0.3533 \\
eMACSJ0840.2+4421-g08   & 08 40 07.77 \,\,$+$44 22 23.8 &  0.6370 & eMACSJ0840.2+4421-g24   & 08 40 11.56 \,\,$+$44 19 09.7 &  0.6490 \\
eMACSJ0840.2+4421-g09   & 08 40 12.96 \,\,$+$44 21 41.6 &  0.6459 & eMACSJ0840.2+4421-g25   & 08 40 14.01 \,\,$+$44 19 33.0 &  0.6402 \\
eMACSJ0840.2+4421-g10   & 08 40 01.53 \,\,$+$44 22 35.5 &  0.6448 & eMACSJ0840.2+4421-g26   & 08 40 11.08 \,\,$+$44 20 05.2 &  0.6427 \\
eMACSJ0840.2+4421-g11   & 08 40 00.24 \,\,$+$44 23 13.4 &  0.6424 & eMACSJ0840.2+4421-g27   & 08 40 15.40 \,\,$+$44 20 16.9 &  0.6432 \\
eMACSJ0840.2+4421-g12   & 08 40 07.22 \,\,$+$44 22 14.1 &  0.6408 & eMACSJ0840.2+4421-g28   & 08 40 11.47 \,\,$+$44 20 17.1 &  0.6305 \\
eMACSJ0840.2+4421-g13   & 08 40 13.66 \,\,$+$44 23 54.0 &  0.6322 & eMACSJ0840.2+4421-g29   & 08 40 16.65 \,\,$+$44 18 33.0 &  0.6410 \\
eMACSJ0840.2+4421-g14   & 08 40 01.90 \,\,$+$44 23 40.5 &  0.6423 & eMACSJ0840.2+4421-g30   & 08 40 14.87 \,\,$+$44 21 02.1 &  0.6359 \\
eMACSJ0840.2+4421-g15   & 08 40 01.19 \,\,$+$44 23 59.9 &  0.6285 & eMACSJ0840.2+4421-g31   & 08 40 22.11  \,\,$+$44 19 14.4 &  0.6485 \\
eMACSJ0840.2+4421-g16   & 08 40 00.84 \,\,$+$44 24 07.1 &  0.6282 & eMACSJ0840.2+4421-g32   & 08 40 07.29 \,\,$+$44 22 41.8 &  0.6333 \\ \hline
eMACSJ0840.2+4421           &        $n_z=30$ & $z= 0.6393$ & \multicolumn{2}{c}{$\sigma = 1310^{+125}_{-205}$} & \\ 

\\ \hline
RXJ0959.0+0255-g01   &       09 59 02.72 \,\,$+$02 54 29.0 &  0.3304 & RXJ0959.0+0255-g23   &       09 59 22.85 \,\,$+$02 53 10.1 &  0.3471 \\                        
RXJ0959.0+0255-g02   &       09 59 01.73 \,\,$+$02 53 36.1 &  0.3517 & RXJ0959.0+0255-g24   &       09 58 37.12 \,\,$+$02 57 05.2 &  0.3522 \\                        
RXJ0959.0+0255-g03   &       09 58 41.58 \,\,$+$03 02 02.9 &  0.3322 & RXJ0959.0+0255-g25   &       09 58 33.79 \,\,$+$03 03 00.6 &  0.3319 \\                        
RXJ0959.0+0255-g04   &       09 59 00.48 \,\,$+$02 55 43.8 &  0.3499 & RXJ0959.0+0255-g26   &       09 58 31.80 \,\,$+$02 59 54.7 &  0.2109 \\                        
RXJ0959.0+0255-g05   &       09 59 00.75 \,\,$+$02 56 03.9 &  0.3458 & RXJ0959.0+0255-g27   &       09 58 40.39 \,\,$+$03 00 53.0 &  0.4959 \\                        
RXJ0959.0+0255-g06   &       09 59 07.33 \,\,$+$02 55 14.6 &  0.3478 & RXJ0959.0+0255-g28   &       09 58 41.21 \,\,$+$02 56 40.6 &  0.6107 \\                        
RXJ0959.0+0255-g07   &       09 59 12.32 \,\,$+$02 55 52.3 &  0.3534 & RXJ0959.0+0255-g29   &       09 58 44.68 \,\,$+$02 58 21.7 &  0.7030 \\                        
RXJ0959.0+0255-g08   &       09 58 56.88 \,\,$+$02 56 20.4 &  0.3464 & RXJ0959.0+0255-g30   &       09 58 45.46 \,\,$+$02 56 38.2 &  0.7094 \\                        
RXJ0959.0+0255-g09   &       09 59 13.26 \,\,$+$02 51 17.6 &  0.2297 & RXJ0959.0+0255-g31   &       09 58 50.61 \,\,$+$02 57 40.9 &  0.5082 \\                        
RXJ0959.0+0255-g10   &       09 59 22.14 \,\,$+$02 52 37.8 &  0.3462 & RXJ0959.0+0255-g32   &       09 58 54.75 \,\,$+$02 57 57.5 &  0.6449 \\                        
RXJ0959.0+0255-g11   &       09 59 16.87 \,\,$+$02 52 52.3 &  0.3477 & RXJ0959.0+0255-g33   &       09 58 51.50 \,\,$+$02 56 07.0 &  0.3512 \\                        
RXJ0959.0+0255-g12   &       09 59 25.07 \,\,$+$02 52 59.0 &  0.3468 & RXJ0959.0+0255-g34   &       09 59 05.36 \,\,$+$02 53 45.4 &  0.3493 \\                        
RXJ0959.0+0255-g13   &       09 59 19.04 \,\,$+$02 52 41.4 &  0.3318 & RXJ0959.0+0255-g35   &       09 59 01.69 \,\,$+$02 55 47.9 &  0.3486 \\                        
RXJ0959.0+0255-g14   &       09 58 40.26 \,\,$+$02 57 52.4 &  0.2205 & RXJ0959.0+0255-g36   &       09 59 03.53 \,\,$+$02 55 21.8 &  0.2115 \\                        
RXJ0959.0+0255-g15   &       09 58 31.77 \,\,$+$02 58 58.6 &  0.3334 & RXJ0959.0+0255-g37   &       09 59 18.53 \,\,$+$02 54 02.9 &  0.4945 \\                        
RXJ0959.0+0255-g16   &       09 58 38.21 \,\,$+$02 59 36.1 &  0.4269 & RXJ0959.0+0255-g38   &       09 59 06.36 \,\,$+$02 54 20.4 &  0.3513 \\                        
RXJ0959.0+0255-g17   &       09 58 36.30 \,\,$+$02 58 12.5 &  0.3552 & RXJ0959.0+0255-g39   &       09 59 11.60 \,\,$+$02 55 24.8 &  0.4240 \\                        
RXJ0959.0+0255-g18   &       09 58 50.42 \,\,$+$02 56 05.9 &  0.3504 & RXJ0959.0+0255-g40   &       09 59 07.67 \,\,$+$02 52 22.4 &  0.3513 \\                        
RXJ0959.0+0255-g19   &       09 58 54.84 \,\,$+$02 59 32.8 &  0.3493 & RXJ0959.0+0255-g41   &       09 59 09.58 \,\,$+$02 51 47.2 &  0.3066 \\                        
RXJ0959.0+0255-g20   &       09 58 51.81 \,\,$+$02 59 17.0 &  0.2821 & RXJ0959.0+0255-g42   &       09 58 36.16 \,\,$+$02 57 26.5   & 0.4977  \\    
RXJ0959.0+0255-g21   &       09 58 57.01 \,\,$+$02 57 19.7 &  0.3477 & RXJ0959.0+0255-g09s  &      09 59 13.67 \,\,$+$02 51 12.9 & 0.6099  \\  
RXJ0959.0+0255-g22   &       09 59 21.07 \,\,$+$02 54 40.6 &  0.3478 & & &\\   \hline 
RXJ0959.0+0255           &        $n_z=21$ & $z= 0.3494$ & \multicolumn{2}{c}{$\sigma = 590^{+90}_{-100}$} &\\ 

\\ \hline
eMACSJ1057.5+5759-g01 &   10 57 31.01 \,\,$+$57 59 45.5 &   0.6015  & eMACSJ1057.5+5759-g19 &   10 57 39.63 \,\,$+$57 56 55.1 &   0.5987  \\                   
eMACSJ1057.5+5759-g02 &   10 57 30.48 \,\,$+$58 00 16.5 &   0.5951  & eMACSJ1057.5+5759-g20 &   10 57 16.83 \,\,$+$57 56 42.7 &   0.5994  \\                   
eMACSJ1057.5+5759-g03 &   10 57 29.99 \,\,$+$57 59 15.4 &   0.5920  & eMACSJ1057.5+5759-g21 &   10 57 24.99 \,\,$+$57 55 45.7 &   0.5969  \\                   
eMACSJ1057.5+5759-g04 &   10 57 36.47 \,\,$+$57 59 13.1 &   0.5899  & eMACSJ1057.5+5759-g22 &   10 57 10.78 \,\,$+$57 55 46.6 &   0.5961  \\                   
eMACSJ1057.5+5759-g05 &   10 57 33.82 \,\,$+$57 58 14.8 &   0.5927  & eMACSJ1057.5+5759-g23 &   10 57 08.15 \,\,$+$57 54 23.3 &   0.7634  \\                   
eMACSJ1057.5+5759-g06 &   10 57 19.17 \,\,$+$57 58 12.1 &   0.5998  & eMACSJ1057.5+5759-g24 &   10 57 28.98 \,\,$+$57 52 59.9 &   0.5756  \\                   
eMACSJ1057.5+5759-g07 &   10 57 41.00 \,\,$+$58 00 53.8 &   0.6050  & eMACSJ1057.5+5759-g25 &   10 57 32.59 \,\,$+$57 58 55.4 &   0.5965  \\                   
eMACSJ1057.5+5759-g08 &   10 57 30.90 \,\,$+$58 00 48.4 &   0.5975  & eMACSJ1057.5+5759-g26 &   10 57 32.92 \,\,$+$58 03 44.2 &   0.7464  \\                   
eMACSJ1057.5+5759-g09 &   10 57 27.65 \,\,$+$57 57 13.6 &   0.5958  & eMACSJ1057.5+5759-g27 &   10 57 27.81 \,\,$+$58 04 12.9 &   0.6023  \\                   
eMACSJ1057.5+5759-g10 &   10 57 36.96 \,\,$+$57 56 05.3 &   0.5973  & eMACSJ1057.5+5759-g28 &   10 57 37.85 \,\,$+$58 04 19.2 &   0.6018  \\                   
eMACSJ1057.5+5759-g11 &   10 57 29.90 \,\,$+$58 00 00.8 &   0.5928  & eMACSJ1057.5+5759-g29 &   10 58 03.43 \,\,$+$58 05 14.0 &   0.5991  \\                   
eMACSJ1057.5+5759-g12 &   10 57 32.09 \,\,$+$58 00 28.0 &   0.5960  & eMACSJ1057.5+5759-g30 &   10 57 56.01 \,\,$+$58 06 01.3 &   0.6527  \\                   
eMACSJ1057.5+5759-g13 &   10 57 22.75 \,\,$+$57 58 25.0 &   0.5954  & eMACSJ1057.5+5759-g31 &   10 57 43.52 \,\,$+$58 04 56.4 &   0.6465  \\                   
eMACSJ1057.5+5759-g14 &   10 57 25.90 \,\,$+$57 58 54.0 &   0.5988  & eMACSJ1057.5+5759-g32 &   10 57 32.73 \,\,$+$58 02 09.9 &   0.6035  \\                    
eMACSJ1057.5+5759-g15 &   10 57 34.46 \,\,$+$58 02 18.6 &   0.6328  & eMACSJ1057.5+5759-g01s &                                                    &  0.6051  \\ 
eMACSJ1057.5+5759-g16 &   10 57 22.56 \,\,$+$58 02 31.6 &   0.7453  & eMACSJ1057.5+5759-g12s &                                                    & 1.0993   \\ 
eMACSJ1057.5+5759-g17 &   10 57 12.29 \,\,$+$57 56 38.2 &   0.6328  & eMACSJ1057.5+5759-g14s &                                                    & 0.6089   \\  
eMACSJ1057.5+5759-g18 &   10 57 35.70 \,\,$+$57 56 31.9 &   0.5967  &  & & \\ \hline 
eMACSJ1057.5+5759         &        $n_z=26$ & $z= 0.5978$ & \multicolumn{2}{c}{$\sigma = 860^{+100}_{-170}$} & \\ \end{tabular}
\end{table*}

\begin{table*}
\contcaption{Galaxy redshifts obtained with Keck-II/DEIMOS during our eMACS pilot study. }
\begin{tabular}{lcclcc}
Galaxy  & R.A.\ \& Dec (J2000) & $z$ & Galaxy  & R.A.\ \& Dec (J2000) & $z$ \\ \hline
eMACSJ1419.2+5326-x1    &       14 19 16.79 \,\,$+$53 25 26.4 &  \,\,\,1.0907$^a$ &   eMACSJ1419.2+5326-g22  &      14 19 30.88 \,\,$+$53 26 10.0 &   0.6347  \\                          
eMACSJ1419.2+5326-x2    &       14 19 12.27 \,\,$+$53 26 48.6 &  \,\,\,2.0027$^a$ &   eMACSJ1419.2+5326-g23  &      14 19 35.17 \,\,$+$53 27 29.0 &   0.6361  \\                         
eMACSJ1419.2+5326-g01  &      14 19 12.13 \,\,$+$53 26 11.6 &   0.6380  &   eMACSJ1419.2+5326-g24  &      14 19 32.04 \,\,$+$53 27 38.9 &   0.6359  \\                          
eMACSJ1419.2+5326-g02  &      14 19 10.92 \,\,$+$53 26 20.8 &   0.6468  &   eMACSJ1419.2+5326-g25  &      14 19 38.95 \,\,$+$53 28 40.4 &   0.6415  \\                          
eMACSJ1419.2+5326-g03  &      14 19 20.67 \,\,$+$53 27 23.2 &   0.6303  &   eMACSJ1419.2+5326-g26  &      14 19 33.72 \,\,$+$53 28 36.2 &   0.6450  \\                          
eMACSJ1419.2+5326-g04  &      14 19 10.51 \,\,$+$53 25 18.1 &   0.6391  &   eMACSJ1419.2+5326-g27  &      14 19 38.67 \,\,$+$53 27 45.3 &   0.6374  \\                           
eMACSJ1419.2+5326-g05  &      14 19 14.16 \,\,$+$53 26 27.9 &   0.644\,\,\,  &   eMACSJ1419.2+5326-g28  &      14 19 22.00 \,\,$+$53 28 15.9 &   0.5571  \\                         
eMACSJ1419.2+5326-g06  &      14 19 08.00 \,\,$+$53 25 22.4 &   0.6462  &   eMACSJ1419.2+5326-g29  &      14 19 34.79 \,\,$+$53 29 21.5 &   0.6308  \\                          
eMACSJ1419.2+5326-g08  &      14 19 21.15 \,\,$+$53 26 36.3 &   0.7220  &   eMACSJ1419.2+5326-g30  &      14 19 46.23 \,\,$+$53 29 12.0 &   0.6348  \\                          
eMACSJ1419.2+5326-g09  &      14 19 21.62 \,\,$+$53 25 46.6 &   0.6430  &   eMACSJ1419.2+5326-g31  &      14 19 43.27 \,\,$+$53 29 17.5 &   0.6410  \\                          
eMACSJ1419.2+5326-g10  &      14 19 26.86 \,\,$+$53 24 50.8 &   0.6377  &   eMACSJ1419.2+5326-g32  &      14 19 48.11 \,\,$+$53 27 11.2 &   0.6746  \\                          
eMACSJ1419.2+5326-g11  &      14 19 10.95 \,\,$+$53 25 40.4 &   0.6411  &   eMACSJ1419.2+5326-g33  &      14 19 54.72 \,\,$+$53 30 35.7 &   0.6389  \\                          
eMACSJ1419.2+5326-g12  &      14 18 55.51 \,\,$+$53 25 21.9 &   0.6408  &   eMACSJ1419.2+5326-g34  &      14 19 50.90 \,\,$+$53 30 17.2 &   0.6326  \\                          
eMACSJ1419.2+5326-g13  &      14 18 42.64 \,\,$+$53 25 27.4 &   0.6455  &   eMACSJ1419.2+5326-g35  &      14 19 53.49 \,\,$+$53 30 29.8 &   0.6426  \\                          
eMACSJ1419.2+5326-g14  &      14 18 46.66 \,\,$+$53 25 51.7 &   0.6409  &   eMACSJ1419.2+5326-g36  &      14 19 51.89 \,\,$+$53 30 27.3 &   0.6349  \\                          
eMACSJ1419.2+5326-g15  &      14 18 41.97 \,\,$+$53 25 29.2 &   0.6451  &   eMACSJ1419.2+5326-g37  &      14 19 49.05 \,\,$+$53 30 21.8 &   0.6278  \\                          
eMACSJ1419.2+5326-g16  &      14 18 37.07 \,\,$+$53 27 40.9 &   0.6808  &   eMACSJ1419.2+5326-g38  &      14 19 50.30 \,\,$+$53 30 11.7 &   0.6310  \\                          
eMACSJ1419.2+5326-g17  &      14 18 41.84 \,\,$+$53 23 48.0 &   0.6411  &   eMACSJ1419.2+5326-g39  &      14 18 33.61 \,\,$+$53 23 00.5 &   0.4673  \\
eMACSJ1419.2+5326-g18  &      14 19 04.45 \,\,$+$53 23 59.1 &   0.5743  &   eMACSJ1419.2+5326-g23s &                                                       &0.6334\\            
eMACSJ1419.2+5326-g19  &      14 19 29.10 \,\,$+$53 25 22.6 &   0.6351  &   eMACSJ1419.2+5326-g09s &                                                       &0.1942\\            
eMACSJ1419.2+5326-g20  &      14 19 34.53 \,\,$+$53 25 45.4 &   0.6319  &   eMACSJ1419.2+5326-g06s &                                                       &0.7675\\            
eMACSJ1419.2+5326-g21  &      14 19 33.90 \,\,$+$53 25 11.3 &   0.6779  & & & \\ \hline 
eMACSJ1419.2+5326         &        $n_z=32$ & $z= 0.6384$ & \multicolumn{2}{c}{$\sigma = 1020^{+80}_{-120}$} & \\ 

\\ \hline 
RXJ1613.7+5542-g1          & 16 13 42.25 \,\,$+$55 42 04.6 & \,\,\,0.1071$^\dagger$ &  RXJ1613.7+5542-g7     & 16 13 41.32 \,\,$+$55 43 43.5 & 0.2656  \\
RXJ1613.7+5542-g2          & 16 13 42.09 \,\,$+$55 41 55.7 & 0.3528 &   RXJ1613.7+5542-g7     &  16 13 32.57 \,\,$+$55 43 50.7 & 0.3498  \\
RXJ1613.7+5542-g3          & 16 13 28.28 \,\,$+$55 41 59.7 & 0.3497 &   RXJ1613.7+5542-g8     &  16 13 33.40 \,\,$+$55 43 04.5 & 0.5099  \\
RXJ1613.7+5542-g4          & 16 13 50.58 \,\,$+$55 42 21.5 & 0.3532 &   RXJ1613.7+5542-g1s   &  & 0.3530 \\
RXJ1613.7+5542-g5          & 16 13 32.08 \,\,$+$55 41 14.4 & 0.3498 &   RXJ1613.7+5542-g3s   &  & 0.3498 \\
RXJ1613.7+5542-g6          & 16 13 40.66 \,\,$+$55 41 53.1 & 0.3578 &   RXJ1613.7+5542-g7s   &  & 0.3501 \\ \hline
RXJ1613.7+5542           &        $n_z=9$ & $z= 0.3512$ & \multicolumn{2}{c}{$\sigma = 590$} & \\  \\
\end{tabular}
  \parbox{\textwidth}{$^a$ QSO / type-1AGN (broad emission lines). }
  \end{table*}

\end{document}